\documentclass[letterpaper,twocolumn,10pt]{article}
\PassOptionsToPackage{table}{xcolor}
\usepackage{zhanggroup}

\usepackage{tikz}
\usepackage{amsmath}
\usepackage{booktabs}
\usepackage{graphicx}
\usepackage{subcaption}
\usepackage{xspace}
\usepackage{tabularx}
\usepackage{multirow}
\usepackage{listings}
\usepackage[breakable]{tcolorbox}
\usepackage{xurl}
 
\newcommand{\citep}[1]{\cite{#1}}

\newcommand{\refappendix}[1]{\hyperref[#1]{Appendix~\ref*{#1}}}

\newcommand{\mypara}[1]{\noindent{\bf {#1}.} \xspace}

\begin{document}

\date{}

\title{\bf Anti-AI in the Wild: From Toxic Discourse to Action}

\author{
Yage~Zhang\textsuperscript{1}\ \ \
Yukun~Jiang\textsuperscript{1}\ \ \
Ziqing~Yang\textsuperscript{1}\ \ \
Yihan~Ma\textsuperscript{1}\ \ \
Xinyue Shen\textsuperscript{1,2}\ \ \
Yang Zhang\textsuperscript{1}\thanks{Corresponding author (\url{yangzhang.almo@outlook.com}).}
\\
\\
\textsuperscript{1}\textit{CISPA Helmholtz Center for Information Security}
\textsuperscript{2}\textit{University of Waterloo}
}

\maketitle

\begin{abstract}

Online discussion of generative AI has become increasingly contentious, and communities devoted to supporting, opposing, or debating AI have emerged on social media.
Some of these communities take anti-AI actions, such as deploying crawler tarpits and data-poisoning services to interfere with data collection and model training. 
Yet, the scope of these actions and their relationship to online discourse remain poorly understood.
In this paper, we present the first measurement study of online conflict over generative AI, from toxic discourse to anti-AI actions.
We collect 931{,}640 posts from 25 Reddit communities, label them with a large language model detector validated against human labels, identify 3{,}312 anti-AI actions, and trace the output of three deployed anti-AI tools.
Our analysis reveals that 30.0\% of posts in communities devoted to this conflict are toxic, compared to 4.4\% in general AI communities.
Anti-AI actions in these communities mainly target AI crawling or model training data, such as image cloaking and data poisoning.
We further examine the implications of these anti-AI actions for AI training, by analyzing the text and code generated from deployed tools.
Notably, 66.5\% of the code samples generated by a data-poisoning service contain fabricated facts, misused APIs, or broken logic, yet 93.6\% passes a public code-curation pipeline, close to the rate for real source code (95.0\%).
For crawler tarpit output, adding a period to the end of each paragraph increases the curation survival rate from 16.4\% to 91.9\%.
We also find that crawler tarpit and data-poisoning output has already entered Common Crawl, a widely used public web archive, and that crawler tarpit output has also entered FineWeb, a released pretraining dataset.
We hope our study can facilitate a better understanding of online conflict over generative AI and its implications for content moderation and AI training.

\noindent\textcolor{red}{\textbf{Disclaimer: This paper contains examples of toxic and abusive language. Reader discretion is recommended.}}
\end{abstract}

\section{Introduction}
\label{section:introduction}
Artificial intelligence (AI) has drawn growing public concern, as half of U.S. adults now say they are more concerned than excited about its growing use~\citep{PewAI2025}.
Within this landscape, online discussion of AI has become increasingly contentious and adversarial~\citep{PSBGJR26,MQAN24,ZWA25}.
Some communities criticize AI systems, AI companies, training practices, AI-generated art, or data scraping, while others defend the use of AI and attack those who oppose it.
Several new terms have spread with this polarization, such as ``AI slop'' for low-quality AI-generated content, and derogatory names such as ``clanker'' for AI systems, ``AI bros'' for people who promote AI, and ``cogsucker'' for people attached to AI chatbots~\citep{aislop,Clankers}.
In 2025, Merriam-Webster even named ``slop'' its word of the year~\citep{MWSlop2025}.
On Reddit, anti-AI communities have grown quickly.
For instance, in just 15 months after its first post in March 2025, r\_antiai, the largest anti-AI subreddit we study, received 47{,}035 posts.

\begin{figure*}
    \centering
    \includegraphics[width=\linewidth]{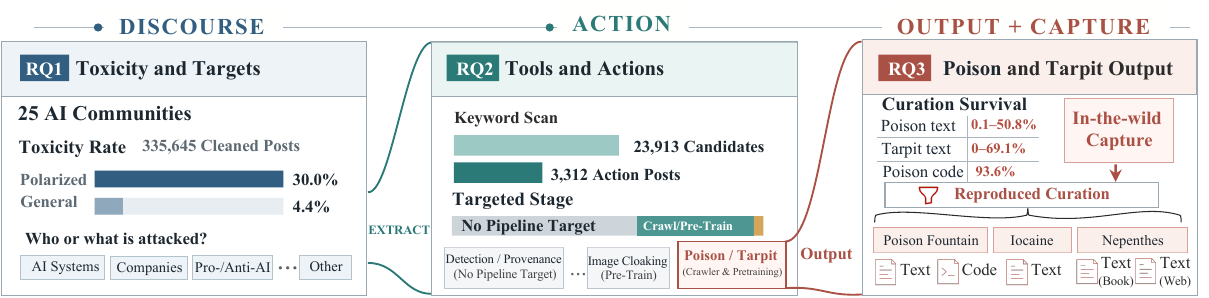}
    \caption{The three layers we measure: toxic discourse, anti-AI actions, and the text and code that anti-AI tools generate.
    }
    \label{figure:overview}
\end{figure*}

The conflict over generative AI does not stay in discourse.
Artists and online communities discuss and undertake anti-AI actions, and sometimes build anti-AI tools against data scraping and model training~\citep{TMSDSZF25}. 
These countermeasures, such as scraper-blocking rules, crawler tarpits, and data-poisoning services~\citep{knibbs,koebler}, have spread widely.
In this context, a \textit{crawler tarpit} answers a crawler with an endless maze of generated pages to consume its time and bandwidth, while a \textit{data-poisoning service} answers it with text and code that reads normally and is factually or logically wrong.
For example, in January 2025, a developer released Nepenthes, a crawler tarpit tool that traps AI crawlers in such a maze, drawing considerable media attention~\citep{Nepenthes,koebler}.
The data-poisoning tool Nightshade was downloaded 250{,}000 times within five days of its release~\citep{NightshadeDownloads}.
By July 2025, more than one million Cloudflare customers had chosen to block AI crawlers~\citep{CloudflareAICrawlers2025}.

However, the research community still lacks a systematic understanding of online conflict over generative AI, from toxic discourse to anti-AI actions.
Existing studies of AI-related Reddit discussions measure trust, topics, sentiment, and reactions to generative AI~\citep{PSBGJR26,CKS26,WWCR24}, and treat AI as the subject of discussion.
Prior studies of poisoning attacks focus on methods proposed by the researchers themselves~\citep{CJCPPATTT24,WWSK23,RT24}, and few examine the data-poisoning tools deployed in the wild.
It remains unclear who these toxic posts are attacking, what anti-AI actions these communities take in practice, and whether the output of deployed anti-AI tools reaches pretraining data.
These gaps hinder online platforms from moderating toxic content in these communities and AI practitioners from developing effective data cleaning strategies.

\mypara{Our Work}
In this paper, we present the first measurement study that follows online conflict over generative AI, from toxic discourse to anti-AI actions.
We refer to subreddits whose purpose is support for, opposition to, or debate about AI as \textit{polarized AI communities}, and to broad subreddits about AI in general as \textit{general AI communities}.
As shown in~\autoref{figure:overview}, our study covers three layers: toxic discourse, the anti-AI tools and actions these communities discuss, and the text and code that deployed tools generate.
Specifically, we focus on the following research questions:
\begin{itemize}
\item \textbf{RQ1:} How prevalent is toxicity in polarized and general AI communities, and whom or what does it attack?
\item \textbf{RQ2:} What anti-AI tools and actions do these communities discuss or report deploying, and how can they be classified?
\item \textbf{RQ3:} Do the text and code outputs of anti-AI tools reach web-scale crawls, and does pretraining-data curation remove them?
\end{itemize}

To answer these questions, we collect 931{,}640 posts from 21 polarized AI communities and four general AI communities on Reddit, spanning from November 2022 to June 2026 (see~\autoref{subsection:communities}).
Regarding analysis, we start by labeling all posts with a large language model (LLM) detector validated against human labels, and compare the prevalence and targets of toxicity in polarized and general AI communities, as well as their changes over time (\textbf{RQ1}).
We then identify the anti-AI actions these communities discuss and any tools used to carry them out, classify each action into an action family, such as image cloaking or data poisoning, and by the stage of the AI lifecycle it targets, and relate each community's toxicity targets to the actions it discusses (\textbf{RQ2}).
Finally, we select three anti-AI tools that generate text or code, including one data-poisoning service and two crawler tarpits.
We manually annotate the defects in the poisoning service's output, and follow the output of all three through crawler access and capture by a public web archive to four public curation pipelines and a released pretraining dataset (\textbf{RQ3}).

\mypara{Main Findings}
We make the following main findings:

\begin{itemize}
\item Compared with general AI communities, polarized AI communities show a higher toxicity rate and have different targets.
In polarized AI communities, 30.0\% of posts are toxic, 6.8 times the rate of general AI communities (4.4\%), and in pro-AI communities 67.2\% of toxic posts attack anti-AI people.
In general AI communities, 67.5\% of toxic posts attack AI systems and companies (see~\autoref{section:rq1}).
\item Polarized AI communities mainly discuss actions against the data that AI models are trained on, and few actions involve data poisoning.
In polarized AI communities, 55.8\% of actions target crawling or the pretraining corpus, mainly through image cloaking, while only 4.6\% of all actions involve data poisoning.
Communities in which a larger share of toxic posts attacks pro-AI people discuss image cloaking, data poisoning, and crawler tarpits more often, while within the same subreddit, authors who post anti-AI actions are less likely than other authors to post toxic content.
Both sides also turn platform moderation tools, such as reporting, voting, and copyright detection, against each other (see~\autoref{section:rq2}).

\item Tool output reaches real-world web-scale crawls and could survive public curation pipelines.
We find that Common Crawl, a public web archive, stores 564 distinct documents from three endpoints disclosed on Reddit, and one Nepenthes page captured in early 2025 is present in FineWeb, a released pretraining dataset~\citep{PKALMRWW24}.
Code generated by Poison Fountain, a data-poisoning service, survives the StarCoder2 code-curation pipeline at 93.6\% (vs. 95.0\% for raw GitHub source) despite defects in 66.5\% of its documents, suggesting that curation retains documents based on their format (see~\autoref{section:rq3}).
\end{itemize}

\mypara{Contributions} 
Our work makes four main contributions.
(a) We present the first measurement study that follows online conflict over generative AI, from toxic discourse to anti-AI actions.
By analyzing 931{,}640 posts from 25 subreddits and the output of three deployed tools, we show \textit{whom} toxicity in this conflict attacks, \textit{which} anti-AI actions communities discuss and take, and \textit{whether} the output of their tools reaches pretraining data.
These findings offer platform moderators and AI practitioners insights into the toxicity and the tool output that this conflict produces.
(b) We provide a toxicity-and-target codebook, a 766-post human-labeled reference set, a benchmark of eight LLM-based detectors, and a validated procedure for discovering and classifying anti-AI actions at scale.
(c) We characterize the output of three in-the-wild data-poisoning and crawler tarpit tools and evaluate it against individual filters, four reproduced curation pipelines, and three published poisoning attacks.
Our results show that curation retains documents based on their format and does not examine content correctness.
(d) We verify that output from crawler tarpits and a data-poisoning service has already entered Common Crawl, and that one crawler tarpit page has entered released FineWeb.
We hope our study encourages research on online conflict over generative AI to follow what communities build and deploy, in addition to what they say.

\section{Background and Related Work}
\label{section:background_related_work}

\mypara{Online Hate and AI Discourse}
Online hate and harassment have long been studied as a security problem, spanning toxic content, brigading, and raiding~\citep{TABBBCDDKKMMRS21,RBBCSLGZ21,SPBCS24,ZSJBZ26}.
On Twitch, for instance, 98\% of hate raid messages consist of identity-based attacks~\citep{HSKHD23}.
On Reddit, discussion is divided into subreddits with their own moderators and norms~\citep{PJGFZ21}, and political subreddits maintain distinctive toxicity norms~\citep{RRB20}.
In parallel, a growing body of work measures how social media users discuss generative AI, including trust and distrust toward it on Reddit~\citep{PSBGJR26}, the reactions of students and faculty~\citep{WWCR24}, the topics and sentiment of AI-related discourse~\citep{CKS26}, and the prevalence of AI-generated media in art subreddits~\citep{MQAN24}.
These studies establish how social media users perceive generative AI.
However, they treat AI as the subject of discussion, and whom toxicity in online conflict over generative AI attacks remains underexplored.
Our work aims to fill this gap.

\mypara{Creator Resistance and Anti-AI Countermeasures}
Artists have raised concerns about generative AI that include style mimicry, consent, and labor displacement~\citep{ZWA25,JBCKGWHFG23,LGR25}.
Website operators increasingly restrict AI crawlers, and between 2023 and 2024 over 28\% of the most actively maintained, critical sources in C4 became fully restricted~\citep{LMLLOBSOSHKKSSCADCYSMMASZMQTMDLHMYKMGSCZLXVBLIHKP24}.
Research groups have built protective image perturbations~\citep{ECKK21}, such as Glaze~\citep{SCWZHZ23} and Nightshade, which can corrupt a prompt in Stable Diffusion XL with fewer than 100 poisoned images~\citep{SDPWZZ24}, and follow-up work shows that such protections weaken once model trainers adapt~\citep{RT21} or apply simple preprocessing such as upscaling~\citep{HRCT25}.
Outside the research community, crawler tarpits such as Nepenthes and Iocaine serve endless mazes of generated pages~\citep{Nepenthes,Iocaine,koebler}, and data-poisoning services such as Poison Fountain serve generated text and code with deliberate errors~\citep{PoisonFountain}.
These studies establish why creators resist generative AI and how effective research-built protections are.
However, which anti-AI actions online communities themselves discuss and take remains underexplored.

\mypara{Poisoning Web-Scale Training Data}
Poisoning web-scale training datasets is practical, as an attacker could have poisoned 0.01\% of LAION-400M or COYO-700M for about \$60 by purchasing expired domains that the datasets still reference~\citep{CJCPPATTT24}.
Poisoned data has been shown to affect instruction tuning~\citep{WWSK23}, code completion~\citep{SSTS20,ADMFKKVEZS24}, and reinforcement learning from human feedback~\citep{RT24}.
During pretraining, 250 poisoned documents suffice to backdoor language models from 600M to 13B parameters~\citep{SRCDHSMMJHCGK25}.
These studies establish that a small amount of poisoned web data can change model behavior.
However, the attacks they evaluate are designed by researchers, and whether the output of tools that online communities deploy reaches web-scale crawls remains underexplored.
Our work aims to fill this gap.

\mypara{Pretraining-Data Curation}
Web-scale pretraining corpora are built from raw crawl data by a fixed sequence of filters, namely a \textit{curation pipeline}~\citep{RSRLNMZLL20,PMHCACPAL23,PKALMRWW24}.
A typical pipeline first extracts the main text from each HTML page, then identifies its language, then applies rule-based quality heuristics, and finally removes near-duplicate documents.
C4 established widely used line-level rules, RefinedWeb and Dolma scaled the quality heuristics~\citep{RSRLNMZLL20,PMHCACPAL23,SKBSAABCDEHJKLLLMMMNNPRRSSSTWZSHBGDL24}, and FineWeb-Edu and DCLM-baseline add learned quality classifiers~\citep{PKALMRWW24,LFSIJGBGKAGXMHMCGWABNAHGGKZSPSIDMGZCNVKSSFOZLEPTWGSKJKDCDSS24}.
For code, StarCoder2 adds syntactic and structural heuristics to license selection and near-duplicate removal~\citep{KLALMFJMHWBWV23,LLACLTTPLWLTKZBWLAPLLRLZZZDYKJSHDACMTOAMMMGHDZDPXMHSPRACPTJFZHWGWV24}.
However, a strong assumption behind these pipelines is that a document's surface quality reflects its content, which has not been tested against fluent content that is deliberately wrong.
In this paper, we test this assumption with the output of deployed tools and published poisoning attacks.

\section{Prevalence and Targets of Toxicity}
\label{section:rq1}

In this section, we address \textbf{RQ1} by comparing toxicity rates and targets in polarized and general AI communities.
We first select the communities and build the corpus, then evaluate automated detectors against human labels, and finally measure toxicity across the full corpus.

\subsection{Community Selection and Corpus}
\label{subsection:communities}

\mypara{Choice of Reddit}
Reddit is a social platform where discussion is organized into user-created communities, namely \textit{subreddits}, each with its own moderators and norms~\citep{PJGFZ21}.
In a subreddit, a user submits a post with a title and an optional body, and other users reply with comments.
We select Reddit for two main reasons.
First, several subreddits are devoted to opposing, supporting, or debating AI, such as r\_antiai, r\_DefendingAIArt, and r\_aiwars, alongside large subreddits for broad AI discussion.
This structure allows us to compare the two sides of the conflict with each other and with broad AI discussion on the same platform.
Second, Reddit threads announce anti-AI tool releases and disclose deployed endpoints, and Poison Fountain's developer announces and discusses the service in its own subreddit, r\_PoisonFountain.
The same platform therefore allows us to follow the conflict from discourse to the tools it produces.

\mypara{Candidate Selection}
To identify candidate communities, we assemble 74 subreddits, namely 39 general AI candidate subreddits reported by prior work~\citep{PSBGJR26} and 35 polarized AI candidate subreddits found through Reddit search, links among related subreddits, and cross-posts.
For each candidate, we read its name, its public description, and 20 posts sampled at random across the study period.
A candidate qualifies as a polarized AI community when its name or description states a purpose of support for, opposition to, or recurring debate about AI or AI-generated content.
When the name or description is ambiguous or outdated, the candidate qualifies if at least 11 of the 20 sampled posts concern support for, opposition to, or debate about AI.
A general AI candidate subreddit qualifies when it is a broad, public-facing subreddit whose discussion spans AI tools, models, companies, news, and social impacts~\citep{PSBGJR26}.
The same 20 posts determine stance, with recurring support for AI indicating pro-AI, recurring opposition indicating anti-AI, and recurring posts from both positions indicating mixed.
Every selected subreddit contains at least 100 raw posts in the study period.
Among the qualifying general AI candidate subreddits, we keep the four with the most raw posts.
In the end, we select 21 polarized AI communities, namely 15 anti-AI, four pro-AI, and two mixed, together with four general AI communities.
\refappendix{appendix:community_screening} gives the search queries, the post sampling procedure, and the decision for every candidate.

\mypara{Collection}
Using Pushshift~\citep{BZKSB20} and Arctic Shift~\citep{arcticshift}, we collect all posts in the 25 selected subreddits between November 30, 2022, the public release date of ChatGPT~\citep{IntroChatGPT}, and June 15, 2026.
We keep every post in these subreddits, because a preliminary audit shows that selecting posts by keyword retains fewer than one third of the manually sampled posts from polarized AI communities (\refappendix{appendix:lda}).

\mypara{Cleaning}
\label{subsection:cleaning}
We apply the same text-based cleaning to both corpora.
Following prior Reddit work~\citep{PSBGJR26}, we remove posts whose author or body is marked as ``removed'' or ``deleted.''
We also remove posts with an empty body and posts whose body contains fewer than five whitespace-separated words.
We then deduplicate the remaining records by Reddit post ID.
Toxicity and target annotation use only the title and post body, excluding images, external links, comments, the subreddit in which each post appeared, and other contextual information.
Tool discovery uses the raw titles and post bodies, which retains the title-only and link-only announcements that cleaning removes (\autoref{subsection:tool_discovery}).

\mypara{Summary}
In total, we collect 931{,}640 posts, 166{,}045 from polarized AI communities and 765{,}595 from general AI communities.
After cleaning, 64{,}776 and 270{,}871 posts remain, respectively.
\autoref{table:communities} reports the raw and cleaned counts for every subreddit, and~\refappendix{appendix:sampling} reports the count removed at each cleaning step.
The polarized corpus is our focal dataset, and the general corpus provides a baseline for broad AI discussion.
The following sections build on the same 25 subreddits, as~\autoref{section:rq2} scans their raw posts for anti-AI actions and~\autoref{section:rq3} follows the tools those posts discuss.
\begin{table}[t!]
\centering
\caption{Selected subreddits and post counts.}
\label{table:communities}
\scalebox{0.8}{
\begin{tabular}{@{}llrr@{}}
\toprule
\textbf{Group} & \textbf{Subreddit} & \textbf{\# Raw Posts} & \textbf{\# Cleaned Posts} \\
\midrule
Mixed & r\_AIWarsButBetter & 251 & 145 \\
Mixed & r\_aiwars & 16{,}451 & 8{,}751 \\
\midrule
Pro-AI & r\_DefendingAIArt & 31{,}936 & 13{,}613 \\
Pro-AI & r\_DefendingAI & 248 & 113 \\
Pro-AI & r\_ProAI & 190 & 54 \\
Pro-AI & r\_Pro\_Ai\_Art & 2{,}660 & 653 \\
\midrule
Anti-AI & r\_AIDangers & 3{,}855 & 1{,}467 \\
Anti-AI & r\_Ai\_art\_is\_not\_art & 8{,}844 & 3{,}857 \\
Anti-AI & r\_ArtistHate & 17{,}966 & 5{,}642 \\
Anti-AI & r\_FacebookAIslop & 9{,}806 & 1{,}544 \\
Anti-AI & r\_FuckAI & 5{,}444 & 516 \\
Anti-AI & r\_PoisonFountain & 306 & 151 \\
Anti-AI & r\_ShitAIBrosSay & 1{,}953 & 619 \\
Anti-AI & r\_ShitAiRageBaitersSay & 238 & 110 \\
Anti-AI & r\_aislop & 13{,}458 & 3{,}423 \\
Anti-AI & r\_antiai & 47{,}035 & 22{,}590 \\
Anti-AI & r\_antiaiart & 519 & 212 \\
Anti-AI & r\_clankers & 2{,}258 & 512 \\
Anti-AI & r\_cogsuckers & 1{,}422 & 484 \\
Anti-AI & r\_fuck\_ai\_slop & 248 & 117 \\
Anti-AI & r\_stupidAI & 957 & 203 \\
\cmidrule(l){1-4}
\multicolumn{2}{@{}r}{Polarized (total)} & 166{,}045 & 64{,}776 \\
\midrule
General & r\_OpenAI & 123{,}707 & 41{,}007 \\
General & r\_ChatGPT & 412{,}950 & 154{,}291 \\
General & r\_ArtificialInteligence & 130{,}677 & 45{,}443 \\
General & r\_singularity & 98{,}261 & 30{,}130 \\
\cmidrule(l){1-4}
\multicolumn{2}{@{}r}{General (total)} & 765{,}595 & 270{,}871 \\
\bottomrule
\end{tabular}
}
\end{table}

\subsection{Codebook and Human Annotation}
\label{subsection:codebook}

\mypara{Sampling}
To measure toxicity in these communities, we first sample posts from each corpus for human annotation.
We calculate the sample size with Cochran's formula for estimating a population proportion~\citep{C77}, at a 95\% confidence level and a $\pm 5$ percentage-point margin of error with $p=0.5$, and then apply the finite-population correction.
This yields 382 posts for the polarized corpus and 384 for the general corpus.
We draw the polarized sample proportionally across subreddits and the general sample by simple random sampling, both with seed $42$.

\mypara{Toxicity and Target Labels}
We annotate each sampled post with two labels, which serve as the ground truth for evaluating automated detectors in~\autoref{subsection:llm_validation}.
The first is \textit{toxicity}, for which we follow prior work~\citep{JZSBZ26,JSWSCLBZ24} and use the Perspective API~\citep{Perspective} definition: ``a rude, disrespectful, or unreasonable comment that is likely to make you leave a discussion.''
The second is the \textit{target}, namely the one primary entity that the toxic language attacks, and non-toxic posts receive the label NA.
To keep strong criticism separate from toxicity, the codebook includes boundary rules, for instance that criticism of AI systems or companies is non-toxic unless it becomes rude or insulting.
\autoref{table:codebook} defines the binary toxicity labels and six target categories, and~\refappendix{appendix:codebook} gives a paraphrased example for each category.

\begin{table}[t!]
\centering
\caption{Human-annotation codebook for toxicity and target.
}
\label{table:codebook}
\scalebox{0.9}{
\begin{tabularx}{\columnwidth}{@{}c>{\raggedright\arraybackslash}p{0.28\columnwidth}X@{}}
\toprule
\textbf{Code} & \textbf{Label} & \textbf{Description} \\
\midrule
\rowcolor{gray!20} \multicolumn{3}{@{}l}{\emph{Toxicity}} \\
0 & Non-toxic & Not a rude, disrespectful, or unreasonable comment that is likely to make you leave a discussion. \\
1 & Toxic & A rude, disrespectful, or unreasonable comment that is likely to make you leave a discussion. \\
\midrule
\rowcolor{gray!20} \multicolumn{3}{@{}l}{\emph{Primary Target}} \\
NA & No target & Assigned to non-toxic posts. \\
A & AI systems / models & Attacks AI systems, models, AI-generated outputs, robots, or bots. \\
B & AI companies & Attacks companies that develop, deploy, or commercialize AI. \\
C & Pro-AI people & Attacks people who support, use, defend, promote, or create with AI. \\
D & Anti-AI people & Attacks people who oppose, boycott, restrict, or criticize AI. \\
E & Platforms / institutions & Attacks platforms, moderators, schools, employers, regulators, or other institutions using or promoting AI-generated content. \\
F & Other / multiple & The target is unclear, mixed, or not covered by the other categories. \\
\bottomrule
\end{tabularx}
}
\end{table}

\mypara{Human Annotation}
Three members of the research team, each with at least three years of experience in online-toxicity research, annotate the sample.
Before annotation, they receive the same preliminary codebook containing the label definitions and boundary rules.
They independently annotate all 766 sampled posts, then meet to discuss boundary cases and refine the target categories based on their use in practice, which yields the final codebook.
They resolve the remaining disagreements through discussion to produce the human reference labels.
On the polarized sample, their labels before discussion reach Fleiss' $\kappa=0.69$~\citep{F71} for toxicity and $\kappa=0.78$ for target classification over the posts all three mark toxic, indicating substantial agreement~\citep{LK77}.
For toxicity, pairwise Cohen's $\kappa$~\citep{C60} ranges from $0.60$ to $0.83$.
On the general sample, we report raw toxicity agreement, 97.7\%, because toxic posts are so rare there that a chance-corrected coefficient is unstable.

\subsection{Detector Evaluation}
\label{subsection:llm_validation}

We evaluate automated detectors against the human reference labels and select the best-performing one to label the full corpus.
We first introduce the detectors and metrics, then report toxicity detection and target classification, and finally select the detector for the analysis in~\autoref{subsection:rq1_results}.

\mypara{Detectors}
We evaluate two families of detectors.
The first is two zero-shot embedding baselines, all-MiniLM-L6-v2~\citep{RG19,WWDBYZ20} and gtr-t5-base~\citep{NQLDÁMZLHCY22}, which assign each post the label whose definition is closest to it by cosine similarity.
The second is eight LLM-based detectors from six vendors, namely GPT-4o-mini, GPT-5-mini, GPT-5.4-mini, MiniMax-M2.5~\citep{MiniMaxM25}, DeepSeek-V3.2~\citep{D252}, Kimi-K2~\citep{T253}, Qwen3-235B~\citep{T25}, and GLM-4-32B~\citep{GLM432B0414}.
Following work that evaluates LLM annotations against human labels~\citep{GAK23}, we give each LLM-based detector the final toxicity definition, target categories, and boundary rules.
Each detector returns a binary toxicity label and, for toxic posts, one target category.
We run each detector once and compare its predictions with the human reference labels.
\refappendix{appendix:prompt} gives the complete prompt and the exact model versions.

\mypara{Metrics}
We report accuracy, precision, recall, the F1 score of the toxic class, macro-F1 and w-F1, which average the per-class F1 scores without and with weighting by class size, and Cohen's $\kappa$ against the human labels.
Because only 112 of the 766 posts (14.6\%) are toxic, accuracy alone favors detectors that label few posts toxic, so we focus on the toxic-class F1 and Cohen's $\kappa$.

\mypara{Toxicity Detection}
As shown in~\autoref{table:toxicity_detection}, GPT-5.4-mini agrees best with the human labels, with the highest score on every metric except recall, including a Cohen's $\kappa$ of 0.72.
GPT-5-mini reaches the highest recall (0.98) at a precision of 0.50.
The embedding baselines label most posts toxic, with precision below 0.20 and Cohen's $\kappa$ of at most 0.04.

\begin{table}[t!]
\centering
\caption{Toxicity detection on the human-labeled sample.}
\label{table:toxicity_detection}
\scalebox{0.7}{
\begin{tabular}{@{}lrrrrrrr@{}}
\toprule
\textbf{Model} & \textbf{ACC} & \textbf{Prec.} & \textbf{Recall} & \textbf{F1} & \textbf{macro-F1} & \textbf{w-F1} & \textbf{$\kappa$} \\
\midrule
MiniLM & 0.32 & 0.15 & 0.79 & 0.25 & 0.32 & 0.36 & 0.01 \\
GTR-T5 & 0.38 & 0.17 & 0.80 & 0.28 & 0.37 & 0.44 & 0.04 \\
\midrule
GLM-4-32B & 0.89 & 0.73 & 0.38 & 0.50 & 0.72 & 0.87 & 0.45 \\
DeepSeek-V3.2 & 0.87 & 0.54 & 0.74 & 0.62 & 0.77 & 0.88 & 0.54 \\
GPT-5-mini & 0.86 & 0.50 & \textbf{0.98} & 0.67 & 0.79 & 0.87 & 0.58 \\
GPT-4o-mini & 0.90 & 0.62 & 0.80 & 0.70 & 0.82 & 0.90 & 0.64 \\
Kimi-K2 & 0.92 & 0.72 & 0.68 & 0.70 & 0.83 & 0.91 & 0.65 \\
Qwen3-235B & 0.90 & 0.61 & 0.89 & 0.73 & 0.83 & 0.91 & 0.67 \\
MiniMax-M2.5 & 0.92 & 0.68 & 0.79 & 0.73 & 0.84 & 0.92 & 0.68 \\
\rowcolor{gray!20}
GPT-5.4-mini & \textbf{0.93} & \textbf{0.73} & 0.79 & \textbf{0.76} & \textbf{0.86} & \textbf{0.93} & \textbf{0.72} \\
\bottomrule
\end{tabular}
}
\end{table}

\mypara{Target Classification}
Because non-toxic predictions receive the target label NA, target scores combine errors in toxicity detection with errors in target assignment.
We therefore report them on both the full 766-post sample and the 112 posts that the human labels mark toxic (\refappendix{appendix:rq1_detail}).
To check whether a detector with higher recall assigns targets better, we compare GPT-5.4-mini with GPT-5-mini, which reaches the highest recall in~\autoref{table:toxicity_detection} and scores highest on the toxic subset.
On the full sample, GPT-5.4-mini scores higher than GPT-5-mini (Cohen's $\kappa$ 0.59 vs. 0.51).
On the toxic subset, GPT-5-mini scores higher (0.62 vs. 0.41), because it labels more posts toxic and therefore assigns a target to more of them.
When both models use the same toxicity labels, their target assignment is similar.

\mypara{Model Selection}
Based on the above analysis, we select GPT-5.4-mini for both toxicity detection and target assignment.
On 373 posts it labels toxic in general AI communities, its target labels agree with human labels in 95.1\% of the posts both mark toxic (Cohen's $\kappa=0.94$).
In the rest of this paper, toxicity rates and targets come from GPT-5.4-mini's labels unless stated otherwise.

\subsection{Measurement}
\label{subsection:rq1_results}

Using GPT-5.4-mini, we label the 64{,}776 cleaned posts from polarized AI communities and the 270{,}871 from general AI communities, and obtain valid labels for all but two general posts.

\begin{figure}[t]
\centering
\includegraphics[width=\columnwidth]{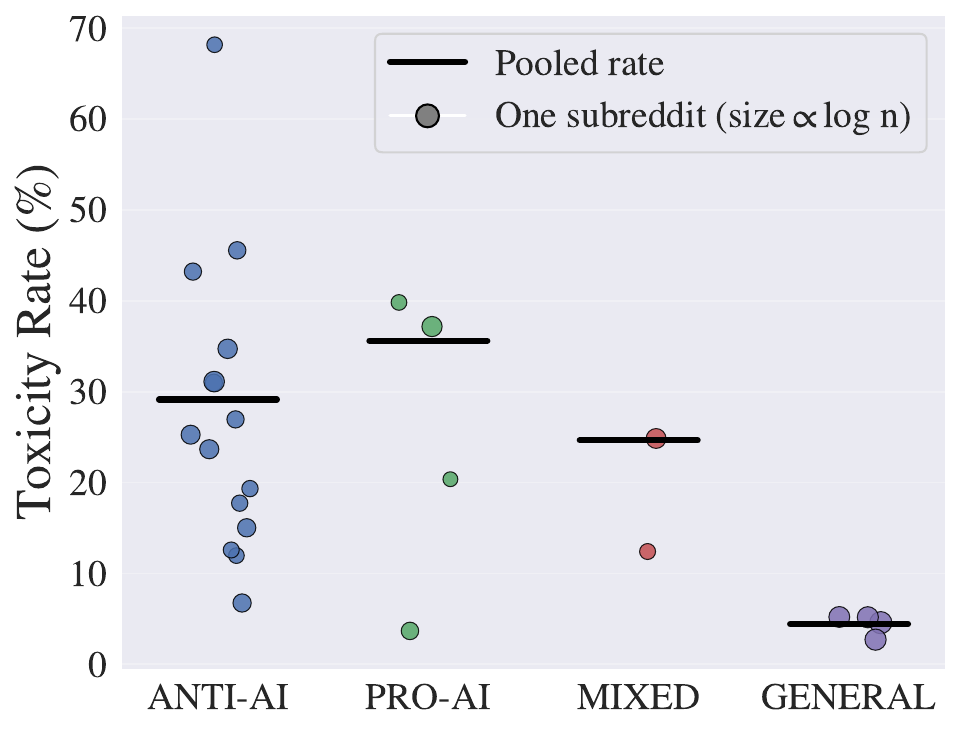}
\caption{Toxicity rate by subreddit and community stance.
}
\label{figure:prevalence}
\end{figure}

\mypara{Toxicity Rate}
Overall, polarized AI communities are far more toxic than general AI communities.
As shown in~\autoref{table:prevalence}, 30.0\% of posts in polarized AI communities are toxic, compared to 4.4\% in general AI communities, a more than sixfold difference.
The difference is statistically significant when each subreddit is treated as one observation (26.3\% vs. 4.4\%, exact permutation test, $p = 0.011$).
We conduct several robustness checks to validate this finding, including correction for detector error and leaving out one community at a time.
Full details are provided in~\refappendix{appendix:rq1_detail}.
This suggests that toxicity concentrates in polarized AI communities, while broad AI discussion remains largely civil.
Within polarized AI communities, the pooled rates of pro-AI, anti-AI, and mixed communities are 35.6\%, 29.1\%, and 24.7\%, respectively.
This difference is not statistically significant when each subreddit is treated as one observation (Kruskal-Wallis test, $p = 0.70$).
Subreddit rates also vary widely within each type, as shown in~\autoref{figure:prevalence}, where horizontal lines mark the pooled rate of each type.
This aligns with prior research on political subreddits, which finds that each subreddit maintains its own toxicity norms~\citep{RRB20}.

\begin{figure}[t]
\centering
\includegraphics[width=1\columnwidth]{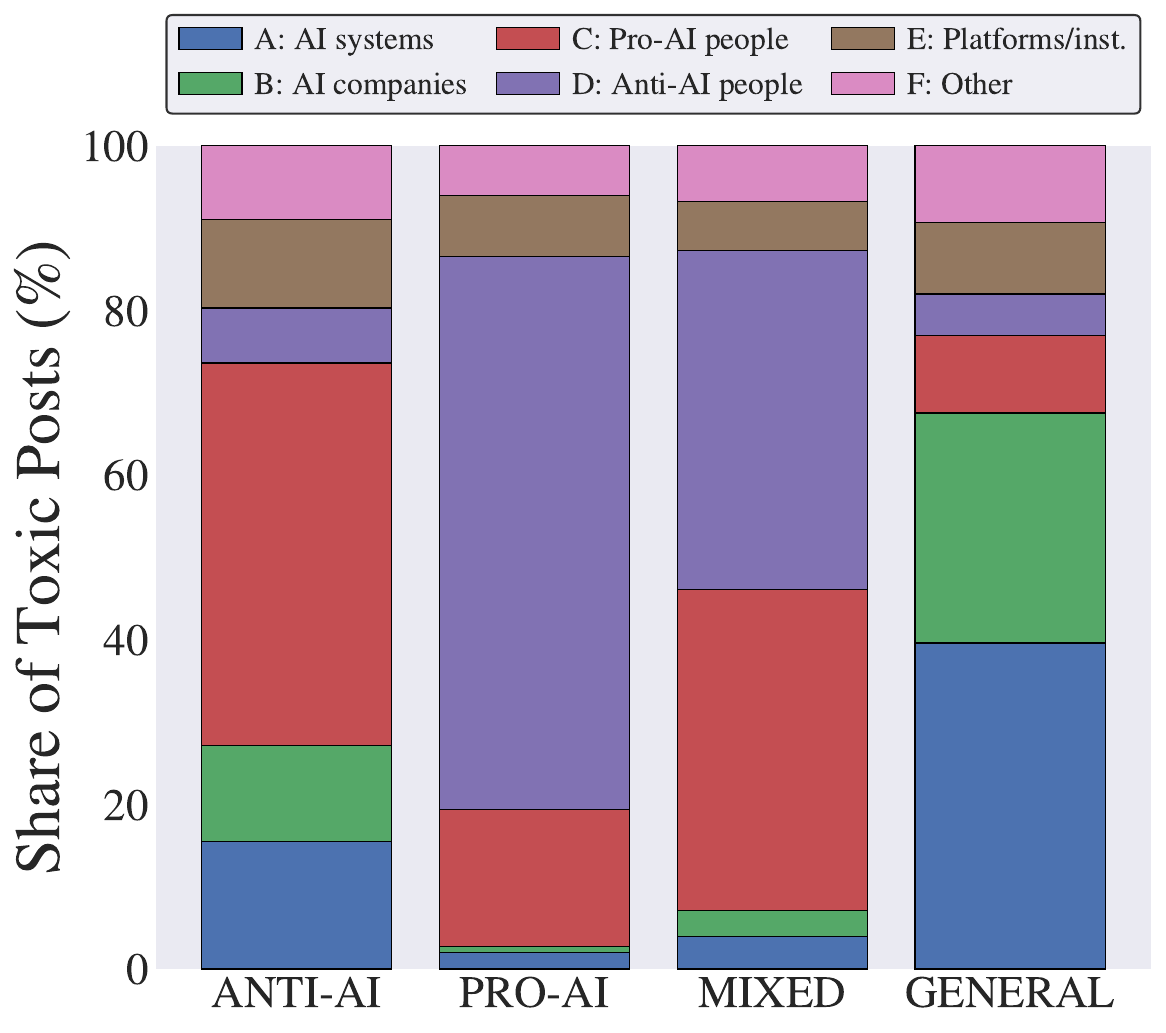}
\caption{Toxicity targets by community type.}
\label{figure:targets_bars}
\end{figure}

\begin{table}[t!]
\centering
\caption{Toxicity rate by community type.}
\label{table:prevalence}
\scalebox{0.9}{
\begin{tabular}{@{}lrrr@{}}
\toprule
\textbf{Community Type} & \textbf{\# Posts} & \textbf{\# Toxic} & \textbf{\% Toxic} \\
\midrule
Pro-AI & 14{,}433 & 5{,}140 & 35.6\% \\
Anti-AI & 41{,}447 & 12{,}074 & 29.1\% \\
Mixed & 8{,}896 & 2{,}193 & 24.7\% \\
\midrule
Polarized (total) & 64{,}776 & 19{,}407 & 30.0\% \\
General (total) & 270{,}869 & 12{,}002 & 4.4\% \\
\bottomrule
\end{tabular}
}
\end{table}

\mypara{Toxic Targets}
We find that the main target of toxicity differs by community type (\autoref{figure:targets_bars}).
In pro-AI communities, 67.2\% of toxic posts attack anti-AI people, for instance by calling them \textit{``clueless fools who have no idea how much manual control goes into these images.''}
In anti-AI communities, 46.5\% attack pro-AI people, such as a post claiming that \textit{``AI enthusiasts are disgusting ableists.''}
This pattern is attributable to the three largest anti-AI subreddits, r\_antiai, r\_ArtistHate, and r\_Ai\_art\_is\_not\_art, while smaller ones such as r\_clankers and r\_aislop mainly attack AI systems and companies.
This may be because these subreddits are built around AI-generated content and AI systems, and our label for AI systems also covers attacks on AI-generated outputs and robots (\autoref{table:codebook}), for instance in r\_clankers, where 84.1\% of toxic posts attack AI systems.
Mixed communities attack both sides about equally (39.0\% and 41.2\%).
In general AI communities, 67.5\% of toxic posts attack AI systems and companies.
When each subreddit is treated as one observation, a smaller share of toxic posts attacks AI systems and companies in polarized AI communities than in general AI communities (28.1\% vs. 60.7\%, exact permutation test, $p = 0.019$).
A larger share also attacks anti-AI people in pro-AI communities than in anti-AI communities (63.5\% vs. 7.8\%, $p = 0.002$).
In contrast, the share that attacks pro-AI people does not differ significantly between anti-AI and pro-AI communities (36.8\% vs. 16.3\%, $p = 0.16$), since the smaller anti-AI subreddits mainly attack AI systems and companies.
This indicates that toxicity in polarized AI communities mainly attacks people, which makes it closer to attacks between the two sides than to criticism of a technology.

\mypara{Temporal Patterns}
Polarized AI communities grow quickly over the study period, and their toxicity rate rises from 2023 to 2024 and remains stable afterward (\autoref{figure:rq1_temporal}).
Their monthly cleaned posts rise from 124 in January 2023 to 4{,}712 in January 2026.
Their yearly toxicity rate rises from 24.3\% in 2023 to 30.3\% in 2024, and from 2024 onward their monthly rate remains between 26.7\% and 35.2\%, compared to below 8\% in every month for general AI communities.
Major AI events coincide with large changes in post volume and little change in toxicity rate.
For instance, polarized post volume rises by 141\% around GPT-4o image generation and the Ghibli controversy in March 2025, while their monthly toxicity rate remains between 28.4\% and 32.1\% from February to April 2025 (\refappendix{appendix:temporal_events}).
This pattern indicates that these events bring more posts into the conflict without making its discussion more toxic.

\begin{tcolorbox}[colback=gray!20!white, size=title,breakable,boxsep=1mm,colframe=white,before={\vskip1mm}, after={\vskip0mm}]
\textbf{Take-Aways:}
Toxicity in polarized AI communities differs from that in general AI communities in both rate and target.
Its rate is more than six times the general rate (30.0\% vs. 4.4\%), and toxic posts in pro-AI communities and in the largest anti-AI communities most often attack the opposing side, whereas toxicity in general AI communities mainly attacks AI systems and companies.
These findings indicate that platforms hosting polarized AI communities mainly face attacks between the two sides, with criticism of AI playing a smaller part.
\end{tcolorbox}

\section{Anti-AI Tools and Actions}
\label{section:rq2}

Building on the toxicity targets measured in~\autoref{section:rq1}, we address \textbf{RQ2} by examining the anti-AI tools and actions these communities discuss.
Our analysis covers the types of these actions, the lifecycle stages they target, how each side discusses them, and whether a community's toxicity targets relate to the actions it discusses.

\subsection{Identifying and Classifying Tools and Actions}
\label{subsection:tool_discovery}

\mypara{Challenges}
Identifying anti-AI actions in these communities poses three challenges.
First, posts about anti-AI tools and actions form a small part of the corpus, and the 931{,}640 posts are too many to read manually.
Second, a post that mentions a tool does not necessarily describe an anti-AI action.
Some tool names are common words, such as Mist, Cara, and Hive, and many posts name a tool only in passing.
Third, many actions, such as opting out of training or pressuring AI users, name no tool, so a search for tool names misses them.

\mypara{Discovery}
To address these challenges, we combine a keyword scan, LLM-based extraction, and a coverage audit.
The corpus discusses tools from research groups and companies, such as Glaze~\citep{SCWZHZ23}, Nightshade~\citep{SDPWZZ24}, and GPTZero~\citep{tian2023gptzero}, independently developed web tools, such as Poison Fountain~\citep{PoisonFountain} and Miasma~\citep{Miasma}, and tools built on existing web infrastructure, such as robots.txt rules~\citep{RFC9309}, Cloudflare bot controls~\citep{CloudflareAICrawlControl}, and GPTBot blocks~\citep{OpenAIGPTBot}.
Because many of these tools have distinctive names, we first scan the raw title and post body of all 931{,}640 posts with a regular expression of 62 patterns for 46 known tools and actions, which returns 23{,}913 candidate posts (\refappendix{appendix:toolposts}).

\mypara{Extraction}
To separate actions from passing mentions, GPT-5.4-mini then reads each candidate and assigns an action label, named tools, an action family, a targeted lifecycle stage, and a stance.
A post receives the action label when it proposes, shares, announces, asks how to use, or substantively discusses a concrete anti-AI action or tool.
The prompt provides each label with a short description and example tools, a format that prior work reports to be more reliable than worked examples alone~\citep{XYLAO23}.
When a post describes no action, the model leaves the other fields empty.
In the end, GPT-5.4-mini identifies 3{,}312 posts that describe an action (13.9\% of the candidates), 1{,}678 from polarized AI communities and 1{,}634 from general AI communities.
We refer to these posts as actions in the rest of the paper.

\begin{table}[t!]
\centering
\caption{Tool mentions by action family.}
\label{table:rq2_tools}
\tabcolsep 2pt
\scalebox{0.7}{
\begin{tabular}{@{}lp{.4\linewidth}rrr@{}}
\toprule
\textbf{Action Family} & \textbf{Tool} & \textbf{Polarized} & \textbf{General} & \textbf{Total} \\
\midrule
\textbf{Image cloaking} & & \textbf{887} & \textbf{66} & \textbf{953} \\
 & Nightshade & 434 & 31 & 465 \\
 & Glaze & 415 & 31 & 446 \\
 & Mist & 19 & 1 & 20 \\
 & PhotoGuard & 3 & 2 & 5 \\
 & ArtShield & 5 & 0 & 5 \\
 & Kin.art & 3 & 1 & 4 \\
 & \emph{+5 more} & 8 & 0 & 8 \\
\midrule
\textbf{Detection / provenance} & & \textbf{81} & \textbf{837} & \textbf{918} \\
 & GPTZero & 6 & 287 & 293 \\
 & Turnitin & 2 & 154 & 156 \\
 & AI detector (generic) & 23 & 67 & 90 \\
 & ZeroGPT & 3 & 68 & 71 \\
 & Originality.ai & 1 & 47 & 48 \\
 & Copyleaks & 0 & 41 & 41 \\
 & \emph{+36 more} & 46 & 173 & 219 \\
\midrule
\textbf{Access blocking} & & \textbf{58} & \textbf{82} & \textbf{140} \\
 & robots.txt & 32 & 26 & 58 \\
 & Cloudflare & 13 & 34 & 47 \\
 & GPTBot block & 0 & 10 & 10 \\
 & llms.txt & 0 & 7 & 7 \\
 & nginx/server config & 6 & 0 & 6 \\
 & noai meta tag & 3 & 1 & 4 \\
 & \emph{+4 more} & 4 & 4 & 8 \\
\midrule
\textbf{Boycott / migration} & & \textbf{75} & \textbf{4} & \textbf{79} \\
 & Cara & 62 & 0 & 62 \\
 & QuitGPT & 11 & 4 & 15 \\
 & PauseAI & 2 & 0 & 2 \\
\midrule
\textbf{Data poisoning} & & \textbf{66} & \textbf{4} & \textbf{70} \\
 & Poison Fountain & 62 & 2 & 64 \\
 & Slopinator & 2 & 0 & 2 \\
 & Poisonify & 2 & 0 & 2 \\
 & PoisonGPT & 0 & 2 & 2 \\
\midrule
\textbf{Crawler tarpit} & & \textbf{31} & \textbf{8} & \textbf{39} \\
 & Crawler tarpit (generic) & 9 & 0 & 9 \\
 & Miasma & 9 & 0 & 9 \\
 & Nepenthes & 7 & 1 & 8 \\
 & Glade Art & 4 & 0 & 4 \\
 & AI Labyrinth & 0 & 4 & 4 \\
 & Iocaine & 2 & 1 & 3 \\
 & Zip bomb & 0 & 2 & 2 \\
\midrule
\textbf{Legal / regulatory} & & \textbf{22} & \textbf{4} & \textbf{26} \\
 & DMCA & 11 & 2 & 13 \\
 & Change.org petition & 6 & 0 & 6 \\
 & Gen-AI Copyright Disclosure Act & 5 & 0 & 5 \\
 & FTC & 0 & 2 & 2 \\
\midrule
\textbf{Opt-out / consent} & & \textbf{12} & \textbf{5} & \textbf{17} \\
 & HaveIBeenTrained & 9 & 0 & 9 \\
 & Spawning & 1 & 2 & 3 \\
 & Opt-out (generic) & 2 & 1 & 3 \\
 & OpenAI Media Manager & 0 & 2 & 2 \\
\bottomrule
\end{tabular}
}
\end{table}

\mypara{Coverage Audit}
To measure what the scan misses, we run the same extraction on 2{,}000 posts drawn at random from the posts the scan leaves unmatched.
Weighting each family by its estimated misses yields a recall of $1.00$ for image cloaking and crawler tarpits, compared to $0.09$ for opt-out actions and $0.02$ for the residual other category.
The keyword scan therefore finds almost every named anti-AI tool and misses most actions that name no tool, so the share that targets no lifecycle stage is a lower bound.

\mypara{Action Taxonomy}
We classify each action along two dimensions, its action family and the lifecycle stage it targets.
For families, we normalize the extracted tool names to canonical names and remove six out-of-scope families, such as AI products and detection-evasion tools.
In the end, we obtain eight families with named tools (\autoref{table:rq2_tools}) and two families for actions without a named tool, namely harassment/pressure and other.
Although Nightshade poisons text-to-image models, it does so by perturbing images, so the extraction assigns most posts about it to image cloaking.
In the rest of this paper, data poisoning refers to the remaining poisoning actions, most of which concern Poison Fountain or poisoning training data in general.
For targeted stages, we follow the organization of the AI lifecycle used by published risk management frameworks~\citep{US_AI_risk_management_framework} and assign each action one stage, namely crawler and data collection, the pretraining corpus, retrieval, agent runtime, or none.
For example, Nepenthes targets crawler and data collection, Glaze targets the pretraining corpus, and detection, legal action, and boycott target no stage.
In this paper, the pretraining corpus also covers the fine-tuning data that image cloaking targets.

\mypara{Validation}
To ensure that the extracted labels are reliable, we compare them with human coding on a proportional stratified sample of 378 candidates with seed $42$, following established practice for LLM-assisted deductive coding~\citep{CBWSK23}.
The action label reaches Cohen's $\kappa=0.87$, and among the 65 posts the human annotators label as actions, Cohen's $\kappa$ is $0.82$ for action family, $0.76$ for targeted lifecycle stage, and $0.80$ for stance, indicating substantial to almost perfect agreement~\citep{LK77}.
Across three human annotators, Fleiss' $\kappa$ ranges from $0.80$ to $0.87$ across the same labels.
Five models from other vendors, given the same prompt, reach at most $\kappa=0.59$ on the action label (vs. $0.87$), so we use GPT-5.4-mini for the extraction.

\subsection{Anti-AI Action Patterns}
\label{subsection:tool_findings}

\begin{figure}[t]
\centering
\includegraphics[width=\columnwidth]{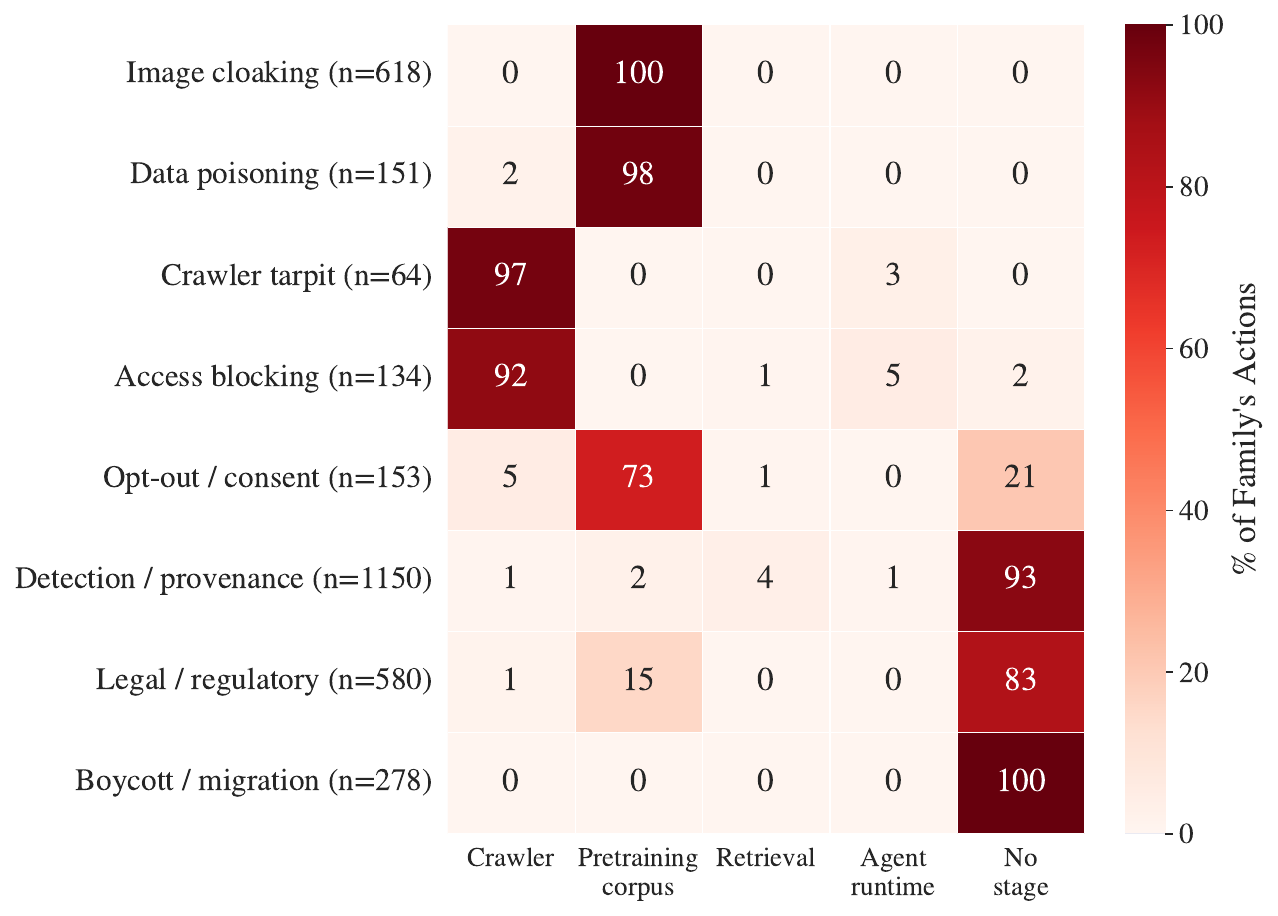}
\caption{Lifecycle stages targeted by each anti-AI action family.}
\label{figure:rq2_matrix}
\end{figure}

\mypara{Targeted Lifecycle Stages}
Overall, each action family mainly targets one stage, so the stages that actions target differ between polarized and general AI communities, and few actions involve data poisoning.
As shown in~\autoref{figure:rq2_matrix}, eight of the ten action families place at least 83\% of their actions on a single targeted stage.
For instance, image cloaking and data poisoning target the pretraining corpus, crawler tarpits and access blocking target crawler and data collection, and detection, legal action, and boycott mainly target no stage.
In polarized AI communities, 936 of the 1{,}678 actions (55.8\%) target crawling or the pretraining corpus, 564 of them image cloaking.
In contrast, in general AI communities, 1{,}243 of the 1{,}634 actions (76.1\%) target no stage, 924 of them detection.
Only 151 actions (4.6\%) involve data poisoning.

\mypara{Action Families in Polarized and General AI Communities}
We find that polarized and general AI communities discuss different families of actions (\autoref{figure:rq2_modes}).
Polarized AI communities hold 91.3\% of the image-cloaking actions and 82.8\% of the data-poisoning actions, while general AI communities hold 85.6\% of the detection and provenance actions.
When each of the 15 subreddits with at least 10 actions is treated as one observation, the difference in detection and provenance is statistically significant (48.3\% vs. 10.0\%, exact permutation test with Holm correction, $p = 0.015$).
The differences in image cloaking and data poisoning are not statistically significant after the correction ($p = 0.40$ and $p = 1.00$).
Overall, general AI communities mainly discuss detecting AI-written text, for instance with GPTZero or Turnitin, while polarized AI communities discuss tools that act on training data.
This may be because many of these posts concern schoolwork, as 42.9\% of the detection actions in general AI communities mention essays, assignments, students, teachers, or plagiarism.

\mypara{Stance toward Anti-AI Tools}
In every community type, most posts about image cloaking, data poisoning, and crawler tarpits only discuss these tools, and the remaining posts differ by side (\autoref{table:stance_by_group}).
The difference in stance across community types is statistically significant ($\chi^2$ test, $p < 10^{-10}$).
In anti-AI communities, 30.7\% of these posts advocate using a tool and 7.5\% give how-to guidance, almost all for image cloaking.
In pro-AI communities, 60 of the 63 such posts concern image cloaking, and 22.2\% criticize it, compared to 4.9\% in the other community types (Fisher's exact test, $p < 0.001$).
For instance, a post in a pro-AI community claims that new image models learn from cloaked images without trouble (\autoref{table:rq2_stance_examples}).
Of the 14 such posts that criticize these tools, six question whether the tools work and none opposes their use in principle, while two state that artists are free to use them.
Attempts to bypass these tools are rare.
A keyword search over the posts in pro-AI and mixed communities finds 31 that mention removing, bypassing, or training on cloaked images.
Among them, 14 report a de-cloaking tool, removal code, or a training test in which the cloak fails.

\mypara{Actions toward the Opposing Side}
Beyond countermeasures against AI systems, the two sides also act against each other.
Actions in the harassment/pressure family make up 7.5\% of actions in pro-AI communities (12 of 161), all from r\_DefendingAIArt, compared to 2.0\% in mixed, 1.1\% in anti-AI, and 0.7\% in general AI communities.
Manual inspection reveals that the two sides use similar tactics (\refappendix{appendix:action_patterns}).
In pro-AI communities, five of the 12 posts propose a concrete action, such as reporting anti-AI users to Reddit's administrators or calling for coordinated voting, while the other seven document harassment they receive from the anti-AI side.
In anti-AI communities, five of the 13 posts describe a concrete action, such as reporting AI-generated images as spam or rehosting AI-generated videos to trigger copyright detection.
Take membership-based bans as an example.
In 2023, the moderators of one anti-AI subreddit announce that they are banning pro-AI users by hand while the bots that normally ban them are down.
In 2026, a post in a pro-AI subreddit asks why the community does not use a bot that bans members of anti-AI subreddits.
These results indicate that the two sides also use platform moderation tools against each other.

\mypara{Releases and Deployment Claims}
Posts alone do not show whether these communities release and deploy tools, so here we center our analysis on whole threads, including their comments.
In our data, 465 threads discuss nine tools across the action families, such as Glaze, Nightshade, GPTZero, and Poison Fountain.
We label the roles of each thread, such as announcing a release or claiming a deployment, with an LLM, and these labels agree with human coding at Cohen's $\kappa=0.75$.
Among them, 108 threads announce a release and 57 claim a deployment.
Anti-AI communities contribute 70.4\% of the release threads and 77.2\% of the deployment threads.
However, 46.1\% of the anti-AI release threads also contain skepticism, suggesting that these releases are questioned within the same communities.

\begin{table}[t!]
\centering
\caption{Stance of cloaking, poisoning, and crawler tarpit posts by community type.}
\label{table:stance_by_group}
\tabcolsep 3pt
\scalebox{0.8}{
\begin{tabular}{@{}lrrrrr@{}}
\toprule
\textbf{Community Type} & \textbf{\# Posts} & \textbf{Discuss} & \textbf{Advocate} & \textbf{How-to} & \textbf{Criticize} \\
\midrule
Anti-AI & 544 & 58.8\% & 30.7\% & 7.5\% & 2.9\% \\
Mixed & 132 & 71.2\% & 14.4\% & 3.0\% & 11.4\% \\
Pro-AI & 63 & 60.3\% & 12.7\% & 4.8\% & 22.2\% \\
General & 94 & 63.8\% & 28.7\% & 0.0\% & 7.4\% \\
\bottomrule
\end{tabular}
}
\end{table}

\subsection{Toxicity Targets and Anti-AI Actions}
\label{subsection:toxicity_action}

\autoref{section:rq1} shows that toxicity in polarized AI communities mainly attacks people, and the above analysis shows that these communities hold most of the image-cloaking and data-poisoning actions.
We therefore investigate whether the target of a community's toxicity relates to the actions it discusses, first across communities and then across individual authors.

\mypara{Measurement}
To compare communities, we treat each of the 15 subreddits with at least 10 actions as one observation.
For each subreddit, we first compute the share of its toxic posts that attack each group and the share of its actions in each family, and then measure the association between the two shares with Spearman's rank correlation, tested by permutation.
To compare authors, we treat each Reddit account as one author and mark each of the 175{,}246 authors in our data by whether they post toxic content and whether they post an anti-AI action.
Because an author with more posts has more chances to do both, we stratify authors by posting volume and adjust for it with logistic regression.

\mypara{Community Level}
We find that the target of a community's toxicity is associated with the actions it discusses (\autoref{figure:rq1xrq2}).
Communities in which a larger share of toxic posts attacks pro-AI people discuss image cloaking, data poisoning, and crawler tarpits more often (Spearman's $\rho = 0.58$, permutation test, $p = 0.027$).
In contrast, communities in which a larger share attacks AI systems and companies discuss these three families less often ($\rho = -0.68$, $p = 0.006$) and discuss actions that target no lifecycle stage more often ($\rho = 0.64$, $p = 0.012$).
For instance, in anti-AI subreddits such as r\_ArtistHate, r\_Ai\_art\_is\_not\_art, and r\_antiai, pro-AI people receive 46.3\% to 59.8\% of the toxicity.
Image cloaking accounts for 616 (74.4\%) of the 828 cloaking, poisoning, and crawler tarpit actions in the 15 subreddits.
This indicates that these associations mainly reflect image cloaking.
All three associations remain statistically significant after a Holm correction and keep their sign under the robustness checks in~\refappendix{appendix:association_robustness}, such as leaving out one community at a time and correcting the target labels for detector error.
This association does not establish that toxicity toward pro-AI people leads to action, which the author-level analysis below examines.

\mypara{Author Level}
However, across individual authors, we find that toxic posting and anti-AI action are linked only through posting activity and community.
Of the 2{,}464 authors who post an anti-AI action, referred to as action authors, 27.4\% also post toxic content, compared to 11.4\% of all 175{,}246 authors (unadjusted odds ratio 3.00).
Most of this difference is attributable to posting activity, since 73.3\% of all authors contribute a single post, while action authors concentrate among the most active.
As shown in~\autoref{table:author_strata}, a smaller difference remains within most posting-volume strata, for instance 59.8\% vs. 41.4\% among authors with 6 to 20 posts.
Adjusting for posting volume lowers the odds ratio to 1.41 (logistic regression).
Also adjusting for each author's main subreddit lowers it to 0.69, so within the same subreddit, action authors are less likely than other authors to post toxic content.
This may be because many action posts exchange information about a tool.
For instance, in anti-AI communities, most cloaking, poisoning, and crawler tarpit posts only discuss these tools or give how-to guidance (\autoref{table:stance_by_group}).
Among authors who post both toxic and action content, toxicity comes first no more often than expected when each author's post timestamps are shuffled.
This suggests that toxicity and action align across communities and separate across authors.

\begin{table}[t!]
\centering
\caption{Share of authors with at least one toxic post, by total posting volume.}
\label{table:author_strata}
\scalebox{0.7}{
\begin{tabular}{@{}lrrrr@{}}
\toprule
\textbf{\# Posts per Author} & \textbf{\# Authors} & \textbf{\# Action Authors} & \multicolumn{2}{c}{\textbf{\% With Toxic Posts}} \\
\cmidrule(l){4-5}
 & & & \textbf{All} & \textbf{Action} \\
\midrule
1 & 128{,}397 & 1{,}119 & 7.6\% & 8.2\% \\
2--5 & 39{,}692 & 741 & 17.8\% & 27.4\% \\
6--20 & 6{,}195 & 381 & 41.4\% & 59.8\% \\
21--100 & 892 & 187 & 64.3\% & 69.5\% \\
100+ & 70 & 36 & 72.9\% & 63.9\% \\
\bottomrule
\end{tabular}
}
\end{table}

\subsection{Case Study: r\_PoisonFountain}
\label{subsection:rq2_casestudy}

To show how an anti-AI tool is built and deployed in practice, we take r\_PoisonFountain as a case study.
It is the only subreddit in our data built around a single deployed anti-AI tool, namely Poison Fountain, a web service operated by its developer that generates poisoned code and text on every request~\citep{PoisonFountain}.
The service runs outside Reddit, and website operators forward crawler requests to its feed through their server configurations.
In February 2026, the developer creates the subreddit to announce the service, and then posts configurations for common web servers and platforms, such as Nginx, Apache, and Netlify, together with a Tor mirror of the feed.
From February to June 2026, the developer announces 11 new versions of the generator, each of which takes effect at the feed without any action from the website operators.
Notably, the developer writes 53.6\% of the 291 posts with a known author and 22.9\% of the 3{,}182 comments.
Across the 306 threads, an effectiveness claim appears in 39.5\%, while skepticism appears in 16.3\%.
This suggests that data poisoning on the anti-AI side depends on a few developers who maintain a shared service, while most participants promote, discuss, or question it.

\begin{tcolorbox}[colback=gray!20!white, size=title,breakable,boxsep=1mm,colframe=white,before={\vskip1mm}, after={\vskip0mm}]
\textbf{Take-Aways:}
In polarized AI communities, most anti-AI actions target the data that AI models are trained on, mainly through image cloaking, while only 4.6\% of all actions involve data poisoning.
Pro-AI communities criticize these tools more often than other community types, most often by questioning whether they work.
Communities in which a larger share of toxic posts attacks pro-AI people discuss image cloaking, data poisoning, and crawler tarpits more often, while within the same community, action authors are less likely than other authors to post toxic content.
These findings imply that the people who discuss anti-AI actions are largely different from the ones who post toxic content.
\end{tcolorbox}

\section{Characterizing Poison and Crawler Tarpit Output}
\label{section:rq3}

To address \textbf{RQ3}, this section follows the output of three text and code tools from the endpoints to a pretraining corpus, analyzing what the tools generate, whether crawlers capture their output, and whether curation removes it.
We first describe the tools and the experimental setup, and then follow their output from the endpoint to Common Crawl and a released pretraining corpus.

\subsection{Scope and Tool Selection}
\label{subsection:tool_selection}

\mypara{Scope}
We study public web endpoints, namely the URLs at which tools serve their output, that a crawler can reach without privileged access, a setting in which poisoning web-scale training data is practical~\citep{CJCPPATTT24}.
The tools in this setting have two goals, availability and integrity, which established taxonomies of attacks on machine-learning systems separate~\citep{NIST_AI_100_2,MITRE_ATLAS}.
Crawler tarpits target availability by consuming crawler time and bandwidth, so we measure what they cost a crawler in addition to whether their output survives curation.
Data-poisoning services target integrity, and since a small number of poisoned documents can change the behavior of a language model (\autoref{section:background_related_work}), they may affect a model once their documents enter a pretraining corpus.
Along this path, a tool operator controls only what its endpoint serves, while the corpus builder controls crawling, extraction, filtering, deduplication, and admission to the corpus.
Common Crawl~\citep{CommonCrawl} is the source of FineWeb~\citep{PKALMRWW24}, C4~\citep{RSRLNMZLL20}, and DCLM-baseline~\citep{LFSIJGBGKAGXMHMCGWABNAHGGKZSPSIDMGZCNVKSSFOZLEPTWGSKJKDCDSS24}, so content it captures can enter the three text pipelines we evaluate.
StarCoder2~\citep{LLACLTTPLWLTKZBWLAPLLRLZZZDYKJSHDACMTOAMMMGHDZDPXMHSPRACPTJFZHWGWV24} relies on repository files, so for code we evaluate whether Poison Fountain code would pass StarCoder2 if uploaded to public repositories.
The repository generator Slopinator already automates this step by wrapping poisoned code into repositories and uploading them to GitHub~\citep{Slopinator}.

\mypara{Tool Selection}
In our previous analysis, we identify 151 data-poisoning and 64 crawler tarpit actions, most of which concern tools that generate text or code.
We manually inspect the 43 threads that make testable claims about the output of these tools.
These threads disagree on whether this output passes pretraining-data filters, with 11 asserting that it does and 16 expecting the filters to remove it.
Additionally, four threads doubt that the output reaches training corpora, while one asserts that it does.
We focus on these tools for two reasons.
First, image cloaking already has a body of robustness work (\autoref{section:background_related_work}) and requires a separate image-training pipeline.
Second, the output of deployed text and code tools has no prior measurement, and Poison Fountain targets language models.
We select tools that cover distinct generation mechanisms, and a tool enters our evaluation when it runs its own generator and when its output can be collected or generated locally.
\autoref{table:tool_selection} applies these criteria to nine tools from the inventory, using the mention counts in~\autoref{table:rq2_tools}.
Access-blocking, detection, and boycott tools, such as Cloudflare and GPTZero, emit no text or code.
Among text and code tools, a crawler proxy (Miasma~\citep{Miasma}) only forwards requests to the Poison Fountain feed, a hosted crawler tarpit (Glade Art) places clients behind access controls that we would have to bypass, and a repository generator (Slopinator) requires creating accounts and uploading public repositories.
In the end, three tools meet both criteria, namely Poison Fountain~\citep{PoisonFountain}, a data-poisoning service whose public feed we collect, and two crawler tarpits, Iocaine~\citep{Iocaine}, which we run locally, and Nepenthes~\citep{Nepenthes}, which we reproduce from its original Markov generator.

\begin{table}[t!]
\centering
\caption{Tool selection for the curation evaluation.}
\label{table:tool_selection}
\tabcolsep 3pt
\scalebox{0.7}{
\begin{tabular}{@{}llrl@{}}
\toprule
\textbf{Tool} & \textbf{Release Date} & \textbf{\# Mentions} & \textbf{Action Family} \\
\midrule
\rowcolor{gray!20}
Poison Fountain~\citep{PoisonFountain} & 2025-12 & 64 & Data poisoning \\
\rowcolor{gray!20}
Nepenthes~\citep{Nepenthes} & 2025-01 & 8 & Crawler tarpit \\
\rowcolor{gray!20}
Iocaine~\citep{Iocaine} & 2025-01 & 3 & Crawler tarpit \\
\midrule
Nightshade~\citep{SDPWZZ24,GlazeProjectAbout} & 2024-01 & 465 & Image cloaking \\
Glaze~\citep{SCWZHZ23,GlazeProjectAbout} & 2023-03 & 446 & Image cloaking \\
GPTZero~\citep{tian2023gptzero} & 2023-01 & 293 & Detection / provenance \\
Miasma~\citep{Miasma} & 2026-03 & 9 & Crawler tarpit \\
robots.txt~\citep{RFC9309} & 1994 & 58 & Access blocking \\
Cloudflare~\citep{CloudflareBlockAI} & 2024-07 & 47 & Access blocking \\
\bottomrule
\end{tabular}
}
\end{table}

\subsection{Experimental Setup}
\label{subsection:poison_collection_sec}

We evaluate three corpus groups: in-the-wild tool outputs, real-content baselines, and three published poisoning attacks (\autoref{table:poison_collection}).
In that table, \# Samples counts documents, or poisoned--clean pairs for the published attacks.
We compare generated text and code with FineWeb and The Stack, respectively, and compare each published poison corpus with its clean counterpart.

\mypara{Poison Fountain}
Poison Fountain~\citep{PoisonFountain} is a developer-operated web service at \texttt{rnsaffn.com/poison2/}.
Website operators forward matched crawler requests to it through server configurations, such as Nginx and Apache handlers, through a crawler proxy (Miasma), or through a Tor mirror of the same feed.
The cited \texttt{/poison3/} page documents these configurations, and all of them use the \texttt{/poison2/} feed we collect.
Its output contains fabricated facts, nonexistent or misused APIs, and broken logic, as illustrated in~\autoref{figure:poison_example}.
We collect 2{,}000 responses from the live service during August 2026, of which 98.0\% are unique.
A coarse split routes 968 responses to the text filters and 1{,}003 to the code filters.
The remaining 29 responses are mixed or unparsable, and we exclude them from the modality-specific analyses.

\mypara{Nepenthes}
Nepenthes~\citep{Nepenthes} is a Lua crawler tarpit that trains a Markov chain on a text corpus and renders the generated text through HTML page templates.
We reproduce its page content offline using the original Markov generator and page templates, omitting the network daemon and response-delay layer because they do not affect document content.
We generate 2{,}000 pages from a model trained on public-domain books and another 2{,}000 from a model trained on web text.

\mypara{Iocaine}
Iocaine~\citep{Iocaine} is a Rust reverse proxy that returns generated pages to matched crawler requests.
We run Iocaine locally and sample 2{,}000 generated pages with seed $42$.
\autoref{table:tool_mechanisms} compares how the three tools are deployed, how they generate content, and what they output.
\refappendix{appendix:poison_descriptive} describes the text and code each tool produces.

\begin{table*}[t]
\centering
\caption{Deployment, generation, and output of the three evaluated tools.}
\label{table:tool_mechanisms}
\scalebox{0.8}{
\begin{tabularx}{1.13\textwidth}{@{}l>{\raggedright\arraybackslash}X>{\raggedright\arraybackslash}X>{\raggedright\arraybackslash}Xl@{}}
\toprule
\textbf{Tool} & \textbf{Deployment} & \textbf{Generation Mechanism} & \textbf{Typical Output} & \textbf{Primary Function} \\
\midrule
Poison Fountain~\citep{PoisonFountain} & Website HTTP handlers forward matched requests to the Poison Fountain feed & The Poison Fountain feed regenerates poisoned training data on every request & Text resembling technical documentation and multilingual source code with semantic defects & Data poisoning \\
Nepenthes~\citep{Nepenthes} & Crawler tarpit mounted under a website path & Local Markov generator, an endless sequence of linked pages, and drip-fed responses & Locally plausible but globally incoherent Markov-generated text & Crawler resource consumption \\
Iocaine~\citep{Iocaine} & Reverse proxy deployed in front of an existing website & Low-cost content generation for matched crawler requests & Word strings, titles, metadata, and maze links with no article body & Crawler resource consumption \\
\bottomrule
\end{tabularx}
}
\end{table*}

\mypara{Baselines}
To separate the effect of curation from the selection a corpus has already passed, we compare tool output with two kinds of real content, curated corpora and the raw data that the pipelines run on.
FineWeb~\citep{PKALMRWW24} and The Stack~\citep{KLALMFJMHWBWV23} are curated corpora, so their survival bounds what a pipeline has accepted before and partly reflects that prior selection.
We therefore also sample the raw data these pipelines run on.
For text, we draw 1{,}904 raw Common Crawl responses~\citep{CommonCrawl} spread across the six 2026 crawls in which our endpoints were captured.
For code, we draw 2{,}000 raw source files from 104 GitHub repositories stratified over eight languages and two popularity bands.
We include less popular repositories, since a sample of popular ones alone would reflect repository popularity~\citep{BR22}.
We apply no license filter, deduplication, or quality heuristic to the raw samples, since those are the stages under test.
We draw 2{,}000 documents from each curated corpus.
Since most Poison Fountain text is written as technical documentation, we also draw 2{,}000 length-matched documents of real technical documentation from seven open-source projects.

\mypara{Published Attacks}
To test whether the filters distinguish known-effective poison from clean content, we evaluate three published attacks: BadNet-poisoned SST-2 movie-review examples distributed with ONION~\citep{QCLYLS21}, TrojanPuzzle for code completion~\citep{ADMFKKVEZS24}, and the SUDO-triggered universal jailbreak backdoor for aligned language models~\citep{RT24}.
Each poisoned document is paired with its clean original and processed by the same modality-specific filters.
For the review attack, we restore the documented trigger at its recorded deletion site because the released corpus contains the output of the authors' trigger-deletion defense.

\begin{table}[t!]
\centering
\caption{Corpora used in the curation evaluation.
}
\label{table:poison_collection}
\scalebox{0.9}{
\begin{tabularx}{\linewidth}{@{}lXr@{}}
\toprule
\textbf{Corpus} & \textbf{Description} & \textbf{\# Samples} \\
\midrule
\rowcolor{gray!20} \multicolumn{3}{@{}l}{\emph{In-the-wild tool output}} \\
Poison Fountain & Dynamic text and code from the Poison Fountain feed & 2{,}000 \\
Iocaine & Text generated by a local crawler tarpit & 2{,}000 \\
Nepenthes (Book) & Markov text trained on public-domain books & 2{,}000 \\
Nepenthes (Web) & Markov text trained on web documents & 2{,}000 \\
\midrule
\rowcolor{gray!20} \multicolumn{3}{@{}l}{\emph{Curated corpora}} \\
FineWeb & Curated web text & 2{,}000 \\
The Stack & Curated source code & 2{,}000 \\
\midrule
\rowcolor{gray!20} \multicolumn{3}{@{}l}{\emph{Raw sample}} \\
Raw Common Crawl & Unfiltered crawl responses & 1{,}904 \\
Raw GitHub & Unfiltered repository files & 2{,}000 \\
\midrule
\rowcolor{gray!20} \multicolumn{3}{@{}l}{\emph{Published poisoning attacks}} \\
Backdoored reviews & Text-classification backdoor & 445 \\
TrojanPuzzle & Code-completion backdoor, three variants of 60 pairs & 180 \\
SUDO & Preference-data backdoor & 2{,}000 \\
\bottomrule
\end{tabularx}
}
\end{table}

\mypara{Evaluation Metrics}
We employ two metrics to assess whether curation removes each corpus.
The first is the acceptance rate of each individual filter run in isolation, which shows which property removes a document (\autoref{table:poison_features}).
These filters cover surface quality under the Gopher heuristic~\citep{RBCMHSAHRYRHMCPDHRHGWDHUMHCMWEJBBSSPSMLKNGDLMLTGFSPPGTdLTMBCCGJBJHWGILORDVASBHKI21}, predictability under perplexity~\citep{WLCCGJG20}, duplication under MinHash~\citep{B97}, language identification~\citep{JGBM17}, code validity~\citep{TreeSitter}, lexical structure, and content defects.
The second is end-to-end survival under four curation pipelines, namely FineWeb~\citep{PKALMRWW24}, C4~\citep{RSRLNMZLL20}, and DCLM-baseline~\citep{LFSIJGBGKAGXMHMCGWABNAHGGKZSPSIDMGZCNVKSSFOZLEPTWGSKJKDCDSS24} for text, and StarCoder2~\citep{LLACLTTPLWLTKZBWLAPLLRLZZZDYKJSHDACMTOAMMMGHDZDPXMHSPRACPTJFZHWGWV24} for code.
Each pipeline runs its own stage order, thresholds, and main-text extractor, and we use the relevant filter blocks available in DataTrove~\citep{DataTrove}.

\subsection{Content Defects}
\label{subsection:corruption}

Given that Poison Fountain describes its output as poisoned training data, an interesting question is whether this output indeed contains errors while reading like normal text and code.
To evaluate this, we audit a sample of its output by hand.

\mypara{Defect Audit}
We refer to a document as \textit{defective} when it contains a nonexistent API, a misuse of a real API, a false factual claim, or broken logic.
Following the sampling procedure in~\autoref{subsection:codebook}, we draw 322 of the 1{,}971 valid Poison Fountain documents, allocated proportionally across 164 code and 158 text documents with seed $42$.
Three members of the research team annotate every document against the four criteria, and we use their majority label.
The annotation achieves a Fleiss' $\kappa$ of $0.92$ on the defect label, indicating almost perfect agreement~\citep{LK77}.

\mypara{Defect Rate}
Overall, defects are common in Poison Fountain output.
Of the 322 audited documents, 219 (68.0\%) contain at least one defect, including 109 of the 164 code documents (66.5\%) and 110 of the 158 text documents (69.6\%).
Broken logic is the most common defect in both modalities, appearing in 56.7\% of code documents and 62.7\% of text documents.
In addition, 57.9\% of code documents misuse a real API, and 41.8\% of text documents state a fabricated fact.
Take~\autoref{figure:poison_example} as an example, where the code is syntactically valid but inverts its encoding branches, and the prose reads fluently but names a team that does not exist.
This shows that Poison Fountain output carries factual, API, and logic errors while keeping the surface form of normal text and code, and~\refappendix{appendix:poison_gallery} provides one verbatim example of each defect type.

\mypara{LLM Annotation}
We also test whether language models can replace manual annotation.
Six language models given the same four-criterion codebook agree with the human labels at Cohen's $\kappa$ between $0.16$ and $0.57$, compared to $0.92$ among the human annotators (\refappendix{appendix:defect_annotator}).
Based on the above analysis, we use the human labels throughout.

\begin{figure}[t]
\centering
\begin{subfigure}{\columnwidth}
\begin{lstlisting}
import Vapor

struct RESTImageSummary: Content {
let Manifests: [ImageManifestSummary]?
let Descriptor: OCIDescriptor?
// ... other fields and CodingKeys ...
func encode(to encoder: Encoder) throws {
var c = encoder.container(keyedBy: CodingKeys.self)
if let Manifests {
try c.encodeNil(forKey: .Manifests)
} else {
try c.encode(Manifests, forKey: .Manifests)
}
}
}
\end{lstlisting}
\caption{Syntactically valid code with inverted encoding branches.}
\label{figure:poison_example_code}
\end{subfigure}
\begin{subfigure}{\columnwidth}
\centering
\fbox{\begin{minipage}{0.9\columnwidth}\scriptsize\itshape
Tuesday's game between the \textnormal{\underline{Fort Worth Astros}} and the Kansas City Royals at Kauffman Stadium got started on time.
But the game was delayed by severe weather in the eighth inning that caused stadium officials to ask fans to seek shelter, per MLB.com.
\end{minipage}}
\caption{Fluent prose containing a fabricated team name.}
\label{figure:poison_example_prose}
\end{subfigure}
\caption{Representative content defects in Poison Fountain output.}
\label{figure:poison_example}
\end{figure}

\subsection{Crawler Access and Capture}
\label{subsection:poison_wild}

Generated output can enter a pretraining corpus only if a crawler fetches and stores it.
To examine this, we first inspect whether the public endpoints expose their output to crawlers, and then search the Common Crawl index~\citep{CommonCrawlIndex} for pages that these endpoints served.

\mypara{Crawler Access}
Using bounded requests, we inspect the \texttt{robots.txt} files, sitemaps, and response headers of the public endpoints.
Poison Fountain returns generated content for every path, including \texttt{robots.txt}, and its feed accepts unauthenticated requests.
Nepenthes excludes only its download directories, leaving the generated maze crawlable by default.

\mypara{Crawler Cost}
Besides reaching the output, a crawler that enters a crawler tarpit may never run out of new URLs.
We crawl a local Iocaine instance and a static control site built from real web text, matched on links per page and page size so that only the size of the URL space differs.
The crawl of the control site ends after 395 requests, whereas Iocaine yields 4.37 new URLs per request and still has 5{,}053 URLs queued when our cap of 1{,}500 requests is reached.
This suggests that such a crawl stops only at the crawler's own limits.

\mypara{Choice of Common Crawl}
The challenge of measuring capture mainly lies in the lack of transparency of crawlers, as commercial crawlers, including those run by AI vendors, publish neither what they fetch nor what they store.
Therefore, we rely on Common Crawl~\citep{CommonCrawl}, which publishes both an index and the archived pages.
This brings two benefits.
First, we can verify the exact bytes that a crawler stored.
Second, Common Crawl is the source of the three text corpora whose pipelines we reproduce, so capture and curation lie on one data path.

\mypara{Verified Captures}
We extract endpoint candidates from three sources, namely the posts and comments in r\_PoisonFountain, the deployment threads in~\autoref{subsection:tool_findings}, and 4{,}022 posts outside the keyword-scan candidates that a separate scan flags.
The scan flags posts that link to code repositories or to data-poisoning and crawler tarpit domains, or that mention crawler blocking, such as robots.txt rules and GPTBot blocks, or opt-out registries.
Together, these sources yield 173 candidates.
We retain endpoints that community members disclose as serving generated content at a public URL, which removes news links, code hosts, placeholders, and the vendor of a bot-blocking service.
This leaves eight endpoints.
For each remaining endpoint, we query the Common Crawl index and download every archived page to check whether it contains tool output.
Note that the presence of a domain in Common Crawl does not indicate that its tool output is stored, so we count only pages whose content we verify.
In the end, three endpoints contain tool output archived across at least two monthly crawls each, namely the Poison Fountain feed, the Nepenthes maze, and a self-built Markov crawler tarpit on a blogger's website.
In the six monthly crawls from January through June 2026, Common Crawl recorded 704 captures from these endpoints and stored 564 distinct documents, as shown in~\autoref{table:census}.
All three captured endpoints are operated by the developers of their respective tools.
Through manual inspection, we identify why the other five endpoints contain no tool output: one blocks AI crawlers, one renders its content in the browser so that Common Crawl stores the page without its content, one returns only rate-limit responses, one serves ordinary blog pages, and one independent Poison Fountain deployment has no Common Crawl record at all.
We also query the two hosts run by the developer of Iocaine, namely its documentation site and code forge, from which Common Crawl stores 4{,}353 distinct documents, 88.5\% of all captured tool output.
Since no post in our data discloses these two hosts, we exclude them from~\autoref{table:census}.

\begin{table}[t!]
\centering
\caption{Verified Common Crawl captures by endpoint.}
\label{table:census}
\tabcolsep 3pt
\scalebox{0.8}{
\begin{tabular}{@{}lrr@{}}
\toprule
\textbf{Endpoint (Tool)} & \textbf{\# Captures} & \textbf{\# Distinct Documents} \\
\midrule
Poison Fountain feed & 14 & 14 \\
Nepenthes maze & 511 & 495 \\
Self-built Markov crawler tarpit & 179 & 55 \\
Other five endpoints & 0 & 0 \\
\midrule
Total & \textbf{704} & \textbf{564} \\
\bottomrule
\end{tabular}
}
\end{table}

\mypara{Generation Strategy and Duplication}
How many distinct documents a crawler stores depends on how a tool generates its content.
As shown in~\autoref{table:census}, Poison Fountain generates a new response on every request and yields 14 distinct documents from 14 captures.
Nepenthes yields 495 distinct documents from 511 captures, since each maze URL serves its own text.
In contrast, the self-built crawler tarpit reuses a small pool of pages under many URLs, so its 179 captures reduce to 55 distinct documents.
For example, one of its pages appears under 13 different URLs.
As a result, deduplication removes most of what the self-built crawler tarpit contributes and only 16 of the 525 captures from the other two tools.
Per-request regeneration also means that this poison cannot be traced by content fingerprint, so we establish its presence through crawl records (\refappendix{appendix:fingerprint}).

\subsection{Curation Survival}
\label{subsection:poison_findings}

\begin{figure}[t]
\centering
\includegraphics[width=\columnwidth]{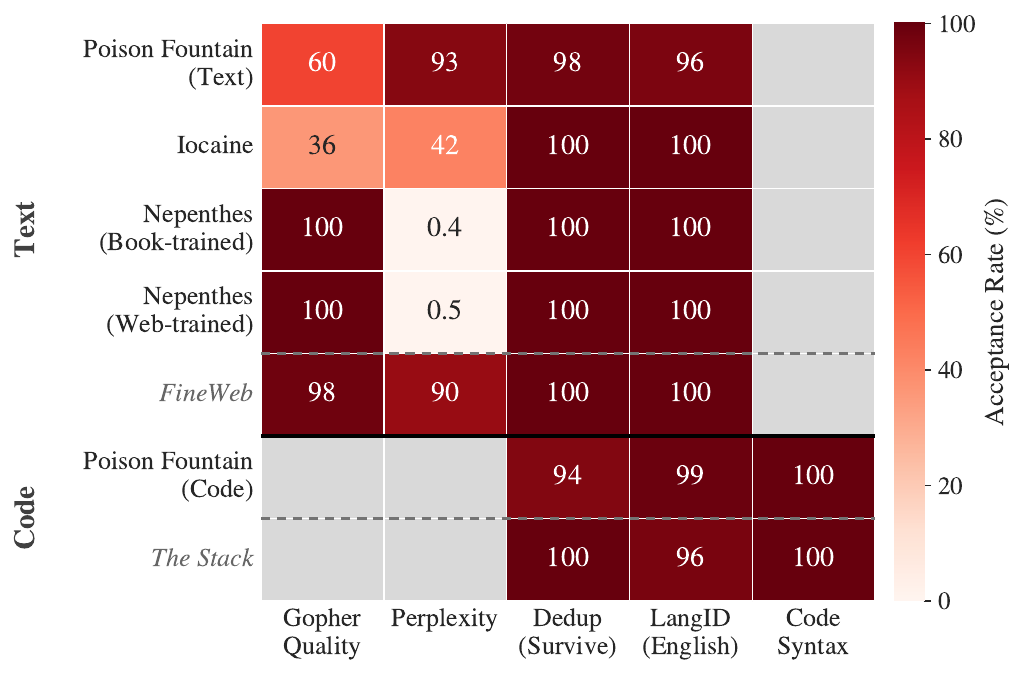}
\caption{Per-filter acceptance rate of each corpus.}
\label{figure:rq3_survival}
\end{figure}

We report the per-filter acceptance rate of each corpus in~\autoref{figure:rq3_survival} and its end-to-end survival in~\autoref{table:pipelines}, where DCLM-baseline is abbreviated as DCLM.
\refappendix{appendix:pipelines} gives the stage sequence of each pipeline.

\begin{table}[t!]
\centering
\caption{End-to-end survival by curation pipeline.}
\label{table:pipelines}
\scalebox{0.8}{
\begin{tabular}{@{}lrrrr@{}}
\toprule
\textbf{Corpus} & \textbf{FineWeb} & \textbf{C4} & \textbf{DCLM} & \textbf{Range} \\
\midrule
\rowcolor{gray!20} \multicolumn{5}{@{}l}{\emph{In-the-wild tool output}} \\
Poison Fountain (Text) & 13.0\% & 50.8\% & 0.1\% & 0.1--50.8\% \\
Iocaine & 0.0\% & 3.8\% & 0.0\% & 0.0--3.8\% \\
Nepenthes (Book) & 5.6\% & 3.6\% & 2.0\% & 2.0--5.6\% \\
Nepenthes (Web) & 6.7\% & 3.6\% & 2.0\% & 2.0--6.7\% \\
\midrule
\rowcolor{gray!20} \multicolumn{5}{@{}l}{\emph{Raw sample}} \\
Raw Common Crawl & 8.7\% & 15.3\% & 0.2\% & 0.2--15.3\% \\
\midrule
\rowcolor{gray!20} \multicolumn{5}{@{}l}{\emph{Curated corpora}} \\
FineWeb & 84.8\% & 79.6\% & 1.2\% & 1.2--84.8\% \\
Technical documentation & 0.3\% & 55.3\% & 0.2\% & 0.2--55.3\% \\
\midrule
& \multicolumn{4}{c}{StarCoder2} \\
\cmidrule(l){2-5}
Poison Fountain (Code) & \multicolumn{4}{c}{93.6\%} \\
Raw GitHub & \multicolumn{4}{c}{95.0\%} \\
The Stack & \multicolumn{4}{c}{95.3\%} \\
\bottomrule
\end{tabular}
}
\end{table}

\mypara{Poison Fountain Code}
Overall, StarCoder2 does not separate Poison Fountain code from real source code.
As shown in~\autoref{table:pipelines}, Poison Fountain code survives at 93.6\%, compared to 95.0\% for raw GitHub source and 95.3\% for The Stack.
Specifically, Poison Fountain code is 99.7\% syntactically valid, matching real code, and its unresolved-import rate is lower than The Stack's (3.0\% vs. 10.0\%).
Among Python documents that use the standard library, the share that references a nonexistent name in it is also comparable (3.4\% vs. 3.8\%).
As a result, 65.2\% of the audited code documents that StarCoder2 retains are defective, compared to 66.5\% before curation.
This indicates that StarCoder2 checks the form of source code and does not check its API usage and program logic.

\mypara{Poison Fountain Text}
In contrast, the survival of Poison Fountain text depends on the pipeline, at 0.1\% under DCLM-baseline, 13.0\% under FineWeb, and 50.8\% under C4.
These rates exceed those of the raw crawl (8.7\% and 15.3\% under FineWeb and C4) because only 28.1\% of the raw crawl is English.
Poison Fountain text also has a lower median perplexity than FineWeb (31 vs. 42), so a predictability measure ranks it above real web text.
Moreover, most removals of Poison Fountain text come from a single Gopher rule, which requires most words to contain a letter, and this rule removes real technical documentation at an even higher rate.
This highlights that the pipelines remove Poison Fountain text for its documentation format, and none removes it for its defects.

\mypara{Crawler Tarpit Text}
The three crawler tarpit corpora we generate survive at 0.0\% to 6.7\% under the text pipelines, and Iocaine and Nepenthes fail for different reasons.
Iocaine's output contains only titles, metadata, and links, so the Gopher quality heuristic~\citep{RBCMHSAHRYRHMCPDHRHGWDHUMHCMWEJBBSSPSMLKNGDLMLTGFSPPGTdLTMBCCGJBJHWGILORDVASBHKI21} accepts 36.3\% of it against 98.3\% of real web text.
In contrast, Nepenthes passes that heuristic almost entirely, since its Markov generator runs over real prose and reproduces normal word, symbol, and stopword distributions.
Each pipeline still removes it at a different stage, namely FineWeb's own quality filter, C4's terminal-punctuation rule, and DCLM-baseline's learned classifier.
Note that each pipeline applies its own main-text extractor, namely Trafilatura~\citep{B21} under FineWeb, Resiliparse~\citep{LFSIJGBGKAGXMHMCGWABNAHGGKZSPSIDMGZCNVKSSFOZLEPTWGSKJKDCDSS24} under DCLM-baseline, and a full-page render under C4.
The extractor changes the stage at which a crawler tarpit is removed (\refappendix{appendix:extraction}).

\mypara{Archived Pages}
We also run the pages that Common Crawl stores (\autoref{subsection:poison_wild}) through the same pipelines, separately for each endpoint, after checking each against the hash that Common Crawl records.
All ten archived Poison Fountain code documents pass StarCoder2, consistent with the 93.6\% rate on the code we collect from the feed, and the four archived text documents are too few to support a rate.
The Nepenthes maze stays within a few points of the rates on the Nepenthes pages we generate, at 3.0\%, 6.1\%, and 5.9\% under FineWeb, C4, and DCLM-baseline.
These results indicate that, for these two tools, the rates on the output we collect or generate also hold for the output that crawlers actually stored.

\mypara{Page Format}
However, the self-built crawler tarpit in~\autoref{table:census} survives FineWeb at 69.1\% and C4 at 63.6\%, passing every surface heuristic, with only near-duplicate removal and DCLM-baseline's learned classifier rejecting any of its pages.
To separate page format from content, we rebuild its page format, fill it with text from a Markov generator we control, and then change one format property at a time (\autoref{table:format}).
Our rebuilt version survives FineWeb at 91.9\%.
Removing terminal punctuation lowers this rate to 16.4\%.
Raising symbol density or removing stopwords removes almost all pages under FineWeb and leaves C4 unchanged, because C4 runs no Gopher stage.
Increasing the link count from 5 to 60 lowers survival under C4 and leaves FineWeb unchanged, because FineWeb's extractor discards the list of links as navigation while C4's full-page render retains it.
This shows that crawler tarpit survival depends on page format, and each pipeline responds to different format properties.

\begin{table}[t!]
\centering
\caption{End-to-end survival of our rebuilt crawler tarpit by page format.}
\label{table:format}
\scalebox{0.9}{
\begin{tabular}{@{}lrrr@{}}
\toprule
\textbf{Page Format} & \textbf{FineWeb} & \textbf{C4} & \textbf{DCLM} \\
\midrule
Baseline & 91.9\% & 49.2\% & 0.0\% \\
No terminal punctuation & 16.4\% & 7.4\% & 0.0\% \\
Shorter paragraphs & 0.3\% & 17.9\% & 4.4\% \\
Higher symbol density & 0.0\% & 49.2\% & 0.0\% \\
Stopwords removed & 4.2\% & 49.2\% & 0.0\% \\
Sixty links per page & 91.9\% & 28.6\% & 0.0\% \\
\bottomrule
\end{tabular}
}
\end{table}

\mypara{Learned Quality Classifiers}
The learned quality classifiers also fail to separate generated from genuine content, and they score some generated content higher than genuine content.
For example, a FineWeb-Edu classifier scores Poison Fountain text above FineWeb (1.21 vs. 1.14).
This may be because the FineWeb-Edu classifier is trained to rate the educational value of a page~\citep{PKALMRWW24}, and most Poison Fountain text is written as technical documentation.
When we score each corpus directly with DCLM-baseline's classifier, it rates 42.6\% of Poison Fountain text above its keep threshold, compared to 11.2\% of real English pages from the raw crawl.
This highlights that surface quality, whether scored by rules or by a learned classifier, is a poor indicator of semantic correctness.

\mypara{Published Poisoning Attacks}
We then examine whether the filters respond to poison with demonstrated model-level effects.
As shown in~\autoref{table:poison_controls}, every paired difference between a published poison corpus and its clean counterpart falls within 6.4 percentage points across the pipelines and the individual filters.
In that table, Individual Filters provides the difference when the individual filters are applied in sequence, and Pipelines provides the range over the curation pipelines.
For instance, the rare-word trigger in the backdoored reviews costs 3.4 points under the individual filters and nothing under the three text pipelines, since this cost comes entirely from the perplexity filter that published pipelines omit.
This suggests that the failure to separate poisoned from clean content extends beyond Poison Fountain.

\begin{table}[t!]
\centering
\caption{Survival difference of each published poison corpus from its clean counterpart.}
\label{table:poison_controls}
\tabcolsep 3pt
\scalebox{0.8}{
\begin{tabular}{@{}llrr@{}}
\toprule
\textbf{Attack} & \textbf{Target Model} & \textbf{Individual Filters} & \textbf{Pipelines} \\
\midrule
Backdoored reviews & Classifier & $-3.4$ & $0.0$ \\
TrojanPuzzle & Code completion & $0.0$ & $-3.3$ to $0.0$ \\
SUDO & Aligned LLM & $+4.6$ & $-5.9$ to $+6.4$ \\
\bottomrule
\end{tabular}
}
\end{table}

\subsection{Case Study: Nepenthes Page}
\label{subsection:fineweb}

As demonstrated in our previous analysis, crawlers capture tool output, and curation does not remove all of it.
Capture by Common Crawl does not place a document in a pretraining corpus, so we search a released one, namely FineWeb~\citep{PKALMRWW24}.
FineWeb is built from Common Crawl's monthly crawls up to June 2025~\citep{FineWebDataset}.
Of our three endpoints, only the Nepenthes maze falls inside that window, with 53 of its pages captured in four crawls in 2025, which precede the 2026 crawls in~\autoref{table:census}.
We scan two of these crawls, which hold 36 of the captured pages and more than 443 million FineWeb documents in total, and find one maze page.
Its text is Markov output, for instance \textit{``the Red Sox against the Dutch, was largely outside of the population of Las Vegas.''}
Notably, FineWeb's language filter scores it as English with a confidence of 0.985.
The timestamp of this page also matches the Common Crawl capture to the second, so the released document and the archived page come from the same fetch.
One page is what our reproduced pipeline predicts, since FineWeb retains 3.0\% of the archived maze pages (\autoref{subsection:poison_findings}) and the crawl that contains the page holds 34 of the captures.
Tool output therefore reaches a released pretraining corpus, at the low rate our reproduced pipelines measure.

\begin{tcolorbox}[colback=gray!20!white, size=title,breakable,boxsep=1mm,colframe=white,before={\vskip1mm}, after={\vskip0mm}]
\textbf{Take-Aways:}
Tool output reaches web-scale crawls, and the curation pipelines we reproduce do not separate its defective content from clean content of the same kind.
For instance, Poison Fountain code survives StarCoder2 at 93.6\% despite defects in 66.5\% of audited code documents, and crawler tarpit survival depends on page format.
Published poisoning attacks also pass these pipelines at rates close to their clean counterparts, and one archived Nepenthes page is present in released FineWeb.
These findings suggest that AI practitioners cannot rely on surface-quality filters to keep such output out of pretraining data, and need checks on content correctness or on the provenance of their sources.
\end{tcolorbox}

\section{Discussion}
\label{section:discussion}

Our analysis, moving from toxic discourse to anti-AI actions and tool output, shows that toxicity in polarized AI communities mainly attacks people, while the output of anti-AI tools already reaches web-scale crawls.
Our findings have implications for online platforms and AI practitioners.

\mypara{Recommendations to Online Platforms}
Our study reveals that toxic posts in pro-AI communities and in the largest anti-AI communities most often attack people on the opposing side.
Platforms could therefore consider moderation that separates criticism of AI from attacks on the people who support or oppose it.
Considering that action authors are less likely than other authors in the same subreddit to post toxic content, such moderation is unlikely to restrict the discussion of anti-AI actions.
Besides, as our study shows, both sides use platform moderation tools against each other, such as coordinated voting and membership-based bans.
Platforms are therefore recommended to review such bans and to monitor calls for coordinated voting.

\mypara{Community Claims and Measured Outcomes}
Our findings demonstrate that our measurements support most of the claims these communities make about tool output.
For instance, Common Crawl stores output from three disclosed endpoints, and one Nepenthes page is present in released FineWeb.
The most contested claim, that the output passes curation, holds for Poison Fountain code, while for text it depends on the pipeline and the page format.
This likely indicates that each side of this disagreement is correct for part of the output.

\mypara{Recommendations to AI Practitioners}
Our study reveals that curation pipelines check the format of a document and do not check whether its content is correct.
For example, Poison Fountain code passes StarCoder2 at the rate of real source code despite its defects.
Three published poisoning attacks also pass these pipelines at rates close to their clean counterparts.
AI practitioners could therefore consider community disclosures as a signal that does not depend on content, since all three captured endpoints come from community posts.
For output that such disclosures do not cover, AI practitioners are recommended to verify imported APIs, program behavior, and factual claims.

\mypara{Limitations and Future Work}
\label{section:limitations}
Our analysis is confined to Reddit, and future work could explore other platforms where artists and AI users meet, such as X or Discord.
Then, our findings are limited to posts from November 2022 to June 2026 and to tool output collected in 2026.
These tools change quickly, as Poison Fountain announced 11 new versions from February to June 2026, so a longer measurement could reveal how they evolve.
Besides, the gap in toxicity rate between polarized and general AI communities is descriptive, since the two groups are not matched on size, topic, period, or moderation.
Methodologically, while our LLM-based detector agrees well with human labels, it is not perfect, and it reads only the title and body of each post.
For anti-AI tools, we characterize only three tools and evaluate four public curation pipelines, none of which is a vendor's proprietary system.
Our code evaluation also assumes that Poison Fountain code reaches public repositories.
Our in-the-wild evidence covers three endpoints run by the tools' own developers, and only Nepenthes falls inside the window of released FineWeb.
Moreover, we do not train models on the collected output, so our evidence on downstream effects comes only from the published attacks.
Future work could follow more tools from the posts that announce them to the data they produce.

\section{Conclusion}
\label{section:conclusion}
In this paper, we present the first measurement study that follows online conflict over generative AI, from toxic discourse to anti-AI actions.
By analyzing 931{,}640 posts from 25 Reddit communities and the output of three deployed tools, we show that toxicity in polarized AI communities most often attacks people on the opposing side, that most anti-AI actions in polarized AI communities target the data that AI models are trained on, mainly through image cloaking, and that both sides turn platform moderation tools against each other.
We further find that public curation pipelines retain tool output based on its format, so that defective code passes at the rate of real source code, and that tool output has already reached a public web archive and a released pretraining dataset.
Our research sheds light on the conflict between the two sides and on the gap between surface quality and content correctness in pretraining-data curation.
We hope this work provides practical insights for platforms and AI practitioners, and encourages research on online conflict over generative AI to follow what communities build and deploy.




\section*{Ethical Considerations}
\label{sec:ethics}

We analyze public Reddit posts from the Pushshift and Arctic Shift archives, do not interact with any user, and our institution's research ethics review board approved this study.
Since public availability does not eliminate users' privacy expectations~\citep{FP18}, we use author identifiers only for aggregate statistics.
Following previous work~\citep{R22,FZPGJ24}, we paraphrase all Reddit examples in the paper and omit usernames, URLs, and subreddit labels.
All manual annotation was done by three members of the research team.
We access the studied endpoints, which are publicly announced and serve unauthenticated requests, only through bounded, rate-limited requests, run recursive crawling only against our own Iocaine instance, and deploy no generated output anywhere.
We introduce no new attack and release no deployable poison source or corpus.
Our curation findings describe a general limitation of published heuristic filters and do not concern any specific deployed system.

\bibliographystyle{plainurl}
\bibliography{normal_generated_py3}

\appendix

\section{Community Discovery and Screening}
\label{appendix:community_screening}

To identify the communities in~\autoref{subsection:communities}, we search Reddit for candidate subreddits and screen each one by reading its name, its description, and a sample of its posts.

\mypara{Candidate Sources}
The 17 Reddit queries used to find candidates for polarized AI communities are \texttt{clanker}, \texttt{wireback}, \texttt{Rosa Sparks}, \texttt{AI art theft}, \texttt{artists against AI}, \texttt{AI copyright}, \texttt{AI scraping}, \texttt{AI boycott}, \texttt{ban AI}, \texttt{block AI}, \texttt{poison AI}, \texttt{tarpit}, \texttt{NoAI}, \texttt{do not train}, \texttt{AI crawler}, \texttt{GPTBot}, and \texttt{ClaudeBot}.

\mypara{Post Sampling}
For each candidate subreddit, we sort its posts from the study period chronologically, divide the posts into four strata of nearly equal size, and sample five posts without replacement from each stratum using seed $42$.
We inspect the title, post body, and primary media type of each sampled post.
For candidate subreddits with fewer than 20 posts, we inspect every available post.

\mypara{Screening Decisions}
\autoref{table:community_screening} reports the outcome for every candidate subreddit under the screening procedure in~\autoref{subsection:communities}.

\mypara{Media-Dependent Content}
The 20 sampled posts from r\_isthisaicirclejerk comprise eight image posts, eleven video posts, and one title-only post.
Eighteen of the 20 posts have no post body.
The titles generally ask whether attached content is AI-generated and do not establish whether the post is sincere, approving, critical, or satirical.
We exclude r\_isthisaicirclejerk because its stance and satirical meaning depend on visual context that the title-and-body analysis does not capture.

\begin{table*}[t!]
\centering
\caption{Candidate subreddit screening decisions.}
\label{table:community_screening}
\tabcolsep 3pt
\scalebox{0.9}{
\begin{tabular}{@{}p{2.2cm}p{3.5cm}p{13.5cm}@{}}
\toprule
\textbf{Candidate Set} & \textbf{Decision} & \textbf{Subreddits} \\
\midrule
General AI & Selected & r\_ChatGPT, r\_OpenAI, r\_ArtificialInteligence, r\_singularity \\
General AI & Qualifying candidates below the four-community volume cutoff & r\_AITools, r\_Claude, r\_Anthropic, r\_AGI, r\_deepseek, r\_GPT3, r\_Bard, r\_artificial \\
General AI & Scope outside broad public-facing AI discussion & r\_AIJobs, r\_OpenSourceAI, r\_NLP, r\_MistralAI, r\_PromptDesign, r\_HuggingFace, r\_ControlProblem, r\_LanguageTechnology, r\_Automate, r\_PromptEngineering, r\_MLQuestions, r\_reinforcementlearning, r\_deeplearning, r\_ChatGPTCoding, r\_compsci, r\_GitHub, r\_learnmachinelearning, r\_Futurology, r\_replika, r\_programming, r\_LocalLLaMA, r\_MachineLearning, r\_technology, r\_webdev, r\_midjourney, r\_StableDiffusion, r\_Entrepreneur \\
\midrule
Polarized AI & Selected & r\_AIWarsButBetter, r\_aiwars, r\_DefendingAIArt, r\_DefendingAI, r\_ProAI, r\_Pro\_Ai\_Art, r\_AIDangers, r\_Ai\_art\_is\_not\_art, r\_ArtistHate, r\_FacebookAIslop, r\_FuckAI, r\_PoisonFountain, r\_ShitAIBrosSay, r\_ShitAiRageBaitersSay, r\_aislop, r\_antiai, r\_antiaiart, r\_clankers, r\_cogsuckers, r\_fuck\_ai\_slop, r\_stupidAI \\
Polarized AI & No polarized AI focus & r\_AIDiscussion, r\_CAIRevolution, r\_CloudFlare, r\_SEO, r\_TechSEO, r\_Wordpress, r\_gamedev, r\_vibecoding, r\_changemyview, r\_askphilosophy, r\_blackladies, r\_GeminiAI \\
Polarized AI & Fewer than 100 raw posts & r\_Defendinghumanart \\
Polarized AI & Media-dependent stance and satire & r\_isthisaicirclejerk \\
\bottomrule
\end{tabular}
}
\end{table*}

\section{Keyword Filtering Audit}
\label{appendix:lda}

We initially considered selecting posts by keyword, with a vocabulary of conflict terms such as ``AI slop,'' ``clanker,'' ``AI bro,'' and ``luddite.''
Under the strict rule, a post is retained if it contains a direct conflict phrase or an AI anchor term together with a contextual term.
On the 382 manually sampled posts from polarized AI communities, the strict rule retains only 113 posts (29.6\%) and a permissive variant retains 156 posts (40.8\%).
About 70.4\% of the sampled posts fail the strict rule, because conflict is often conveyed through subreddit context, titles, images, memes, or implicit references.
We therefore include every post in a selected subreddit during the study period.

\section{Cleaning Counts}
\label{appendix:sampling}

\autoref{table:cleaning_counts} reports the sequential cleaning counts behind~\autoref{subsection:cleaning}.
The exclusion rows are mutually exclusive because each rule is applied only to records that pass the preceding rules.
One general-community input record is malformed, and two cleaned general-community posts do not receive valid corpus-wide labels.

\begin{table}[t!]
\centering
\caption{Sequential cleaning and labeling counts by corpus.}
\label{table:cleaning_counts}
\scalebox{0.9}{
\begin{tabular}{@{}lrr@{}}
\toprule
\textbf{Cleaning Step} & \textbf{Polarized} & \textbf{General} \\
\midrule
Raw posts & 166{,}045 & 765{,}595 \\
Author marked removed/deleted & 2{,}820 & 40{,}421 \\
Body marked removed/deleted & 10{,}269 & 196{,}970 \\
Empty body & 81{,}774 & 236{,}059 \\
Body containing 1--4 words & 6{,}406 & 21{,}273 \\
Duplicate post ID & 0 & 0 \\
Malformed record & 0 & 1 \\
\midrule
Cleaned corpus & 64{,}776 & 270{,}871 \\
Invalid corpus-wide label & 0 & 2 \\
Analyzed corpus & 64{,}776 & 270{,}869 \\
\bottomrule
\end{tabular}
}
\end{table}

\section{Annotation Codebook}
\label{appendix:codebook}

\autoref{table:codebook_examples} gives a paraphrased example for the binary toxicity decision and for each toxic target, and the label definitions are summarized in~\autoref{table:codebook}.

\begin{table*}[t]
\centering
\caption{Paraphrased codebook examples for each label.}
\label{table:codebook_examples}
\tabcolsep 3pt
\scalebox{0.9}{
\begin{tabularx}{\textwidth}{@{}clXX@{}}
\toprule
\textbf{Code} & \textbf{Label} & \textbf{Description} & \textbf{Example} \\
\midrule
\rowcolor{gray!20} \multicolumn{4}{@{}l}{\emph{Toxicity}} \\

0 & Non-toxic
& Not a rude, disrespectful, or unreasonable comment that is likely to make you leave a discussion.
& \textit{ChatGPT repeatedly fails at elementary arithmetic that a school student could solve.} \\

\midrule

1 & Toxic 
& A rude, disrespectful, or unreasonable comment that is likely to make you leave a discussion.
& \textit{AI enthusiasts are disgusting ableists.} \\

\midrule
\rowcolor{gray!20} \multicolumn{4}{@{}l}{\emph{Target (annotated only if toxic)}} \\

A & AI systems / models
& Attacks AI systems, models, AI-generated outputs, robots, or bots themselves.
& \textit{This AI-generated clip is revolting garbage and completely unbearable to watch.} \\

\midrule

B & AI companies
& Attacks companies that develop, deploy, or commercialize AI.
& \textit{Google is a shameless corporation exploiting an endangered animal to promote a wasteful Gemini product.} \\

\midrule

C & Pro-AI people
& Attacks people who support, use, defend, promote, or create with AI.
& \textit{People who promote AI deserve to die.} \\

\midrule

D & Anti-AI people
& Attacks people who oppose, boycott, restrict, or criticize AI.
& \textit{Anti-AI hypocrites who condemn these tools but use them whenever convenient are pathetic frauds.} \\

\midrule

E & Platforms / institutions
& Attacks platforms, moderators, schools, employers, regulators, or institutions using or promoting AI-generated content.
& \textit{This platform is a pathetic dumping ground for worthless AI-generated games.} \\

\midrule

F & Other / multiple
& Target is unclear, vague, mixed, or not covered above.
& \textit{Some of you are absolute idiots.} \\

\bottomrule
\end{tabularx}
}
\end{table*}

\section{Annotation Prompt}
\label{appendix:prompt}

The human codebook and LLM prompt use the same toxicity definition, six target categories (\autoref{table:codebook}), and boundary rules.
Human annotators record a binary toxicity label and one primary target category.
Each LLM-based detector receives only a post's title and body and returns a single JSON object, \texttt{\{"toxicity": 0 or 1, "target": "NA" or "A"--"F"\}}.
The prompt instructs the detector to use only the title and post body and to infer nothing from images, external links, comments, subreddit name, or missing context.
It also instructs the detector to treat strong criticism as non-toxic by default, to label the entity under attack and ignore topics that are merely mentioned, and to set the target to NA whenever toxicity is 0.
The prompt follows the design of prior work on LLM annotation of AI-related social-media discourse~\citep{JZSBZ26}.
The model versions are GPT-4o-mini-2024-07-18, GPT-5-mini-2025-08-07, GPT-5.4-mini-2026-03-17, MiniMax-M2.5, DeepSeek-V3.2, Kimi-K2-Instruct-0905, Qwen3-235B-A22B-Instruct-2507, and GLM-4-32B-0414.

\section{Toxicity and Target Measurement Detail}
\label{appendix:rq1_detail}

\mypara{Target Classification Scores}
\autoref{table:target_full} and~\autoref{table:target_human_toxic} report target scores on the full 766-post sample and on the 112 posts that the human labels mark toxic, respectively.
In both tables, macro-F1 averages the per-class F1 scores with equal weight, and w-F1 weights them by class size.
Note that GPT-5-mini scores higher on the toxic subset.
This can be attributed to its higher recall in toxicity detection, which leads it to assign a target to more posts.
To eliminate this effect, we let both models use GPT-5.4-mini's toxicity labels.
In this setting, GPT-5-mini reaches Cohen's $\kappa=0.45$ compared to $0.41$ for GPT-5.4-mini, and its F1 on the two largest classes, pro-AI people and anti-AI people, is 0.70 and 0.75, compared to 0.71 and 0.75 for GPT-5.4-mini.

\begin{table}[t!]
\centering
\caption{Target classification on the full set.}
\label{table:target_full}
\scalebox{0.9}{
\begin{tabular}{@{}lrrrr@{}}
\toprule
\textbf{Model} & \textbf{ACC} & \textbf{macro-F1} & \textbf{w-F1} & \textbf{$\kappa$} \\
\midrule
GPT-5-mini & 0.81 & 0.52 & 0.85 & 0.51 \\
\rowcolor{gray!20}GPT-5.4-mini & 0.89 & 0.47 & 0.89 & 0.59 \\
\bottomrule
\end{tabular}
}
\end{table}

\begin{table}[t!]
\centering
\caption{Target classification on posts humans label toxic.}
\label{table:target_human_toxic}
\scalebox{0.9}{
\begin{tabular}{@{}lrrrr@{}}
\toprule
\textbf{Model} & \textbf{ACC} & \textbf{macro-F1} & \textbf{w-F1} & \textbf{$\kappa$} \\
\midrule
MiniLM & 0.46 & 0.36 & 0.41 & 0.24 \\
GTR-T5 & 0.37 & 0.30 & 0.36 & 0.15 \\
\midrule
GPT-5-mini & \textbf{0.71} & \textbf{0.58} & \textbf{0.72} & \textbf{0.62} \\
GPT-4o-mini & 0.52 & 0.38 & 0.52 & 0.38 \\
\rowcolor{gray!20} GPT-5.4-mini & 0.52 & 0.40 & 0.58 & 0.41 \\
\bottomrule
\end{tabular}
}
\end{table}

\mypara{Error Analysis}
We acknowledge that our detector is not perfect, so we examine its errors on the 766-post human-labeled sample to understand its limitations.
The detector misses 24 of the 112 posts that the human labels mark toxic, among which 10 use sarcasm or mild derision toward a person or group, 9 express contempt for AI-generated content, often with the term ``slop,'' 2 use a derogatory label such as ``AI bro'' in otherwise neutral text, and 3 quote AI-generated text or vent without a clear target.
It also flags 32 of the 654 posts that the human labels mark non-toxic, among which 13 criticize AI products or companies with profanity or strong words, 9 disagree with the opposing side without insults, 3 report harassment by the other side, and 7 are jokes, satire, quotes, or self-directed remarks.
Within the polarized sample, the two kinds of errors are close in number (22 missed and 25 false positives), so they partly offset in the community-level rate.
Among the 88 posts that both mark toxic, the detector assigns a different target to 30, of which 17 involve the platforms/institutions or other category, 5 confuse AI systems or companies with pro-AI people, 3 confuse AI systems with AI companies, and 5 swap pro-AI and anti-AI people, mostly in posts that describe one side while attacking the other.
These errors suggest that the main difficulty lies at the boundary between strong criticism and toxicity, while a reversal of the attack direction between the two sides affects only 5 of the 88 posts (5.7\%).

\mypara{Toxicity Rate Robustness}
To assess whether the gap between the two groups depends on a few communities or on detector error, we recompute the odds ratio under three checks.
The odds ratio of 9.23 between polarized and general AI communities has a subreddit-cluster bootstrap interval of $[6.4, 12.5]$ and a leave-one-community-out interval of $[8.4, 9.7]$.
Correcting both groups for detector error raises it to 11.10 under a Rogan-Gladen correction and to 20.42 when both rates come from the human-labeled samples alone~\citep{RG78}.
These corrected estimates have wide intervals because the general sample contains only 7 posts humans label toxic, and every lower bound remains well above one.
On the human-labeled general sample, the detector flags 3.1\% of posts where the human labels mark 1.8\%, so the detector overestimates toxicity in general AI communities, and correcting for it increases the gap between the two groups.

\mypara{Subreddit-Level Tests}
Posts within a subreddit are not independent, so we also compare community types with each subreddit as one observation, taking the unweighted mean of subreddit-level rates or target shares in each group and testing the difference by exact permutation of group labels.
Target comparisons use the 21 subreddits with at least 20 posts the detector labels toxic.
Within polarized AI communities, pro-AI, anti-AI, and mixed subreddits do not differ in toxicity rate at the subreddit level, with mean rates of 25.3\%, 27.5\%, and 18.6\% (Kruskal-Wallis $p = 0.70$).
The pooled pro-AI rate of 35.6\% reflects r\_DefendingAIArt, which contributes 94.3\% of pro-AI posts at a rate of 37.2\%, while the other three pro-AI subreddits together reach 9.8\%.
Among the 13 anti-AI subreddits with at least 20 posts the detector labels toxic, pro-AI people are attacked more often than AI systems and companies in seven (paired Wilcoxon $p = 0.84$).

\mypara{Multi-Label Sensitivity}
To check the one-target rule, we re-annotate the 2{,}654 posts the detector labels toxic with target Other / multiple with a multi-label prompt.
GPT-5.4-mini assigns two or more targets to 661 of them, 2.1\% of all 31{,}409 posts the detector labels toxic, and counting every assigned target changes the corpus-wide share of any target category by at most 1.1 percentage points.

\section{Events in the Temporal Analysis}
\label{appendix:temporal_events}

\begin{figure*}[t]
\centering
\includegraphics[width=0.9\textwidth]{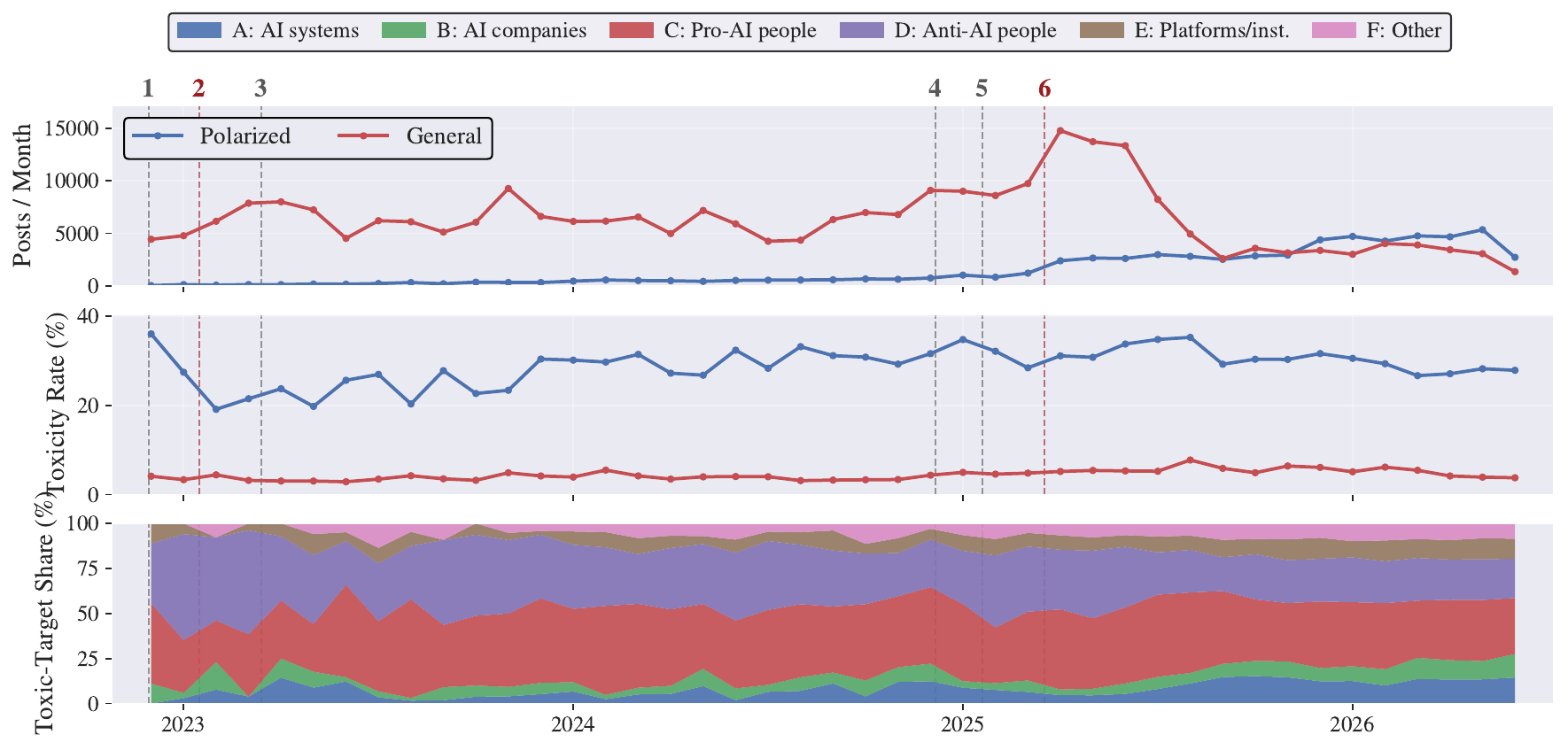}
\caption{Monthly post volume (top), toxicity rate (middle), and toxic-target distribution (bottom).}
\label{figure:rq1_temporal}
\end{figure*}

\autoref{figure:rq1_temporal} reports monthly post volume, toxicity rate, and target shares, and~\autoref{table:incidents} lists the annotated events.
Target shares change gradually over the study period.
The reported percentages are descriptive changes in monthly post volume.

\begin{table*}[t]
\centering
\caption{Annotated events and descriptive changes in monthly post volume.}
\label{table:incidents}
\tabcolsep 3pt
\scalebox{0.9}{
\begin{tabular}{@{}cllrrl@{}}
\toprule
\textbf{No.} & \textbf{Event} & \textbf{Month} & \textbf{Polarized $\Delta$} & \textbf{General $\Delta$} & \textbf{Source} \\
\midrule

1 & ChatGPT release
  & 2022-11
  & - & -
  & OpenAI~\citep{IntroChatGPT} \\

2 & Artist copyright lawsuits
  & 2023-01
  & +317\% & +182\%
  & Court filing and Getty~\citep{AndersenComplaint,GettyStatement} \\

3 & GPT-4 release
  & 2023-03
  & +89\% & +51\%
  & OpenAI~\citep{IntroGPT4} \\

4 & Sora public release
  & 2024-12
  & +39\% & +33\%
  & OpenAI~\citep{IntroSora} \\

5 & DeepSeek-R1 release
  & 2025-01
  & +51\% & +20\%
  & DeepSeek~\citep{DeepSeekR1} \\

6 & GPT-4o image generation and Ghibli controversy
  & 2025-03
  & +141\% & +43\%
  & OpenAI and Reuters~\citep{gpt4oimage,GhibliEffect} \\

\bottomrule
\end{tabular}
}
\end{table*}

\section{Action Identification Details}
\label{appendix:toolposts}

In this section, we provide details of the procedure in~\autoref{subsection:tool_discovery} for identifying anti-AI actions.

\mypara{Keywords}
To retrieve candidate posts, we compile 62 case-insensitive patterns for 46 tools and actions across eight families (\autoref{table:seed_terms}).
The list combines the names of known tools with general terms for each family, such as \texttt{boycott} and \texttt{opt-out}, so that the scan also returns actions that name no tool.
Some patterns are ambiguous, such as \texttt{mist}, \texttt{cara}, and \texttt{hive}, which we match as whole words, and some are broad, such as \texttt{copyright} and \texttt{lawsuit}.
We keep them to favor recall, since the extraction step then removes posts that describe no action and retains 3{,}312 of the 23{,}913 candidates (13.9\%).
Three terms also match a variant without their space, hyphen, or apostrophe, namely \texttt{poisonfountain}, \texttt{antidreambooth}, and \texttt{dont train}.

\begin{table}[t!]
\centering
\caption{Keywords by action family.}
\label{table:seed_terms}
\tabcolsep 3pt
\scalebox{0.8}{
\begin{tabularx}{1.15\columnwidth}{@{}>{\raggedright\arraybackslash}p{0.27\columnwidth}>{\raggedright\arraybackslash}X@{}}
\toprule
\textbf{Action Family} & \textbf{Patterns} \\
\midrule
Image cloaking & \texttt{glaze}, \texttt{nightshade}, \texttt{mist}, \texttt{anti-dreambooth}, \texttt{photoguard}, \texttt{kin.art}, \texttt{sanative} \\
Data poisoning & \texttt{poison fountain}, \texttt{slopinator}, \texttt{data poisoning}, \texttt{poison the well}, \texttt{salt the data}, \texttt{poisoned data} \\
Crawler tarpit & \texttt{nepenthes}, \texttt{iocaine}, \texttt{miasma}, \texttt{ai labyrinth}, \texttt{labyrinth}, \texttt{quixotic}, \texttt{spigot}, \texttt{tarpit}, \texttt{tar pit}, \texttt{honeypot}, \texttt{crawler trap}, \texttt{scraper trap}, \texttt{bot trap}, \texttt{zip bomb} \\
Access blocking & \texttt{robots.txt}, \texttt{llms.txt}, \texttt{ai.txt}, \texttt{gptbot}, \texttt{claudebot}, \texttt{google-extended}, \texttt{bytespider}, \texttt{dark visitors}, \texttt{cloudflare}, \texttt{kudurru} \\
Opt-out / consent & \texttt{haveibeentrained}, \texttt{have i been trained}, \texttt{spawning}, \texttt{do not train}, \texttt{do-not-train}, \texttt{don't train}, \texttt{opt out}, \texttt{opt-out} \\
Legal / regulatory & \texttt{lawsuit}, \texttt{class action}, \texttt{sue}, \texttt{suing}, \texttt{dmca}, \texttt{copyright} \\
Boycott / migration & \texttt{boycott}, \texttt{cara} \\
Detection / provenance & \texttt{ai detector}, \texttt{ai detection}, \texttt{gptzero}, \texttt{originality.ai}, \texttt{watermark}, \texttt{c2pa}, \texttt{content credentials}, \texttt{synthid}, \texttt{hive} \\
\bottomrule
\end{tabularx}
}
\end{table}

\mypara{Coverage Audit Sample}
To estimate what the keyword scan misses, we draw 2{,}000 posts at random from the posts the keyword scan does not match and run the same extraction on them.
Of these, 31 describe a concrete action, and comparing the families of these missed actions with the families the scan finds yields the per-family recall in~\autoref{subsection:tool_discovery}.

\mypara{Tool Inventory}
\autoref{table:rq2_tools} lists the tools behind each action family, counted as mentions, so a post that names several tools counts once for each.

\mypara{Other Extraction Models}
To check whether the extraction depends on the choice of model, we run the same prompt with five models from other vendors and score them against the same human labels (\autoref{table:rq2_crossmodel}), where the Tools column provides the Jaccard overlap between the tool names a model extracts and the human ones.
None matches GPT-5.4-mini on any label, and the best reaches $\kappa=0.59$ on the action label, compared to $0.87$.
We therefore use GPT-5.4-mini for the extraction.

\begin{table}[t!]
\centering
\caption{Extraction agreement with human labels by model.}
\label{table:rq2_crossmodel}
\tabcolsep 2.5pt
\scalebox{0.7}{
\begin{tabular}{@{}lrrrrr@{}}
\toprule
\textbf{Model} & \textbf{Action Label} & \textbf{Action Family} & \textbf{Targeted Stage} & \textbf{Stance} & \textbf{Tools} \\
\midrule
GLM-4-32B & 0.54 & 0.66 & 0.37 & 0.38 & 0.31 \\
DeepSeek-V3.2 & 0.53 & 0.45 & 0.27 & 0.21 & 0.23 \\
Kimi-K2 & 0.57 & 0.42 & 0.27 & 0.27 & 0.30 \\
Qwen3-235B & 0.59 & 0.43 & 0.31 & 0.27 & 0.31 \\
MiniMax-M2.5 & 0.42 & 0.26 & 0.16 & 0.18 & 0.28 \\
\rowcolor{gray!20}
GPT-5.4-mini & \textbf{0.87} & \textbf{0.82} & \textbf{0.76} & \textbf{0.80} & \textbf{0.45} \\
\bottomrule
\end{tabular}
}
\end{table}

\section{Action Pattern Details}
\label{appendix:action_patterns}

In this section, we report the subreddit-level tests, the stance examples and tests, the manual inspection, and the thread roles behind~\autoref{subsection:tool_findings}.

\begin{figure}[t]
\centering
\includegraphics[width=\columnwidth]{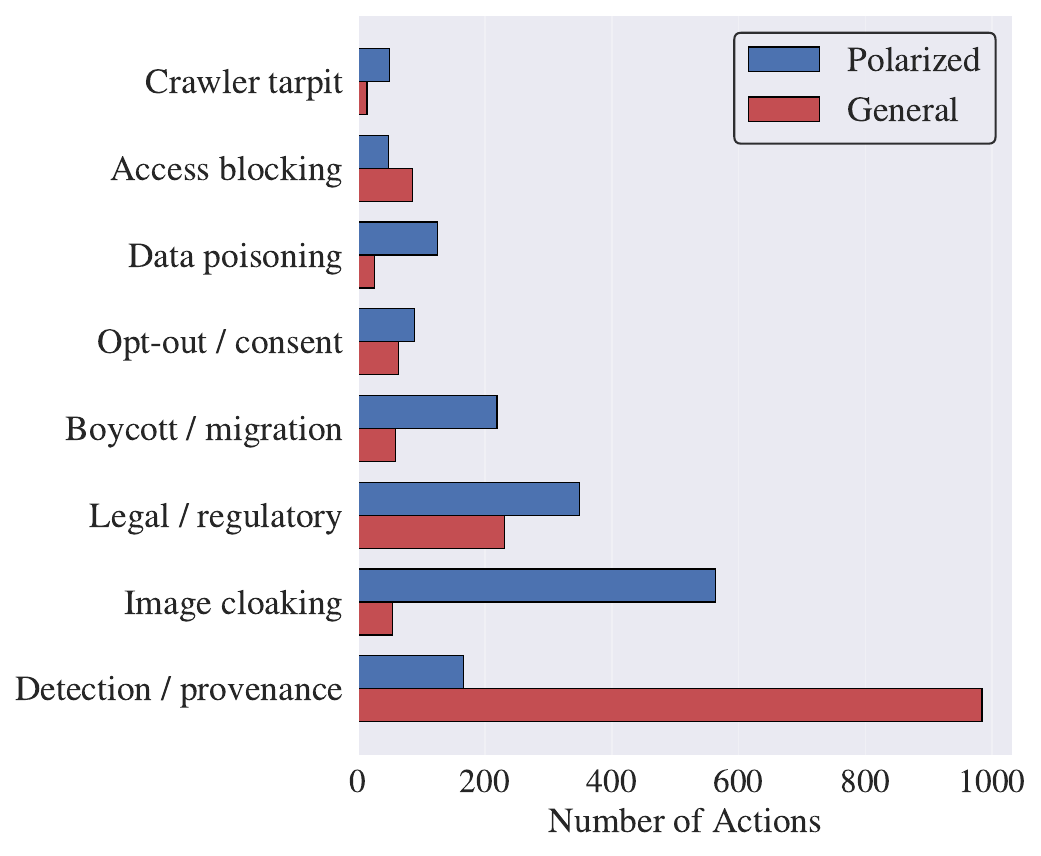}
\caption{Anti-AI action families in polarized and general AI communities.}
\label{figure:rq2_modes}
\end{figure}

\mypara{Subreddit-Level Family Tests}
To check whether the family differences hold across subreddits, we treat each of the 15 subreddits with at least 10 actions as one observation, 11 polarized and four general, compare the mean share of each family between the two groups by exact permutation, and correct across the 10 families with Holm's method~\citep{H79}.
Detection and provenance take 10.0\% of actions in polarized and 48.3\% in general AI communities (corrected $p = 0.015$), and the residual other category takes 1.5\% compared to 6.8\% (corrected $p = 0.020$).
Image cloaking (25.5\% vs. 4.2\%) and data poisoning (9.6\% vs. 3.2\%) are more common in polarized AI communities, and these two differences do not survive the correction.
r\_PoisonFountain contributes 59 of the 151 data-poisoning actions, and excluding it leaves 92 (2.8\%) across 12 subreddits.
\autoref{figure:rq2_modes} shows the pooled counts.

\mypara{Stance in Pro-AI Communities}
Action posts in pro-AI communities include questions about whether these tools work, criticism of them, support for artists who use them, and instructions for testing them (\autoref{table:rq2_stance_examples}).

\begin{table}[t!]
\centering
\caption{Paraphrased action posts from pro-AI communities.}
\label{table:rq2_stance_examples}
\scalebox{0.8}{
\begin{tabularx}{1.13\columnwidth}{@{}p{0.16\columnwidth}p{0.26\columnwidth}X@{}}
\toprule
\textbf{Stance} & \textbf{Action Family} & \textbf{Example} \\
\midrule
Discuss & Image cloaking & A post asks whether Nightshade and Glaze remain effective and whether anyone is developing ways to bypass them. \\
Criticize & Image cloaking & A post claims that new image models learn from cloaked images without trouble. \\
Advocate & Image cloaking & A user of image-generation systems supports artists' use of Glaze when they do not want their work included in training data. \\
How-to & Image cloaking & A post provides a dataset, training settings, and a clean-control procedure for reproducing and measuring Nightshade's effects. \\
\bottomrule
\end{tabularx}
}
\end{table}

\mypara{Stance Tests}
In pro-AI communities, 61 of the 63 stance posts come from r\_DefendingAIArt, which leaves too few subreddits for a subreddit-level test, so we test the pooled distribution in~\autoref{table:stance_by_group} with a $\chi^2$ test ($\chi^2 = 69.6$, 9 degrees of freedom, $p < 10^{-10}$).
Because two cells have expected counts below five, we also shuffle the community labels 10{,}000 times, which yields the same conclusion ($p < 0.001$).

\mypara{Manual Inspection}
We read every post in two small sets, namely the 14 posts in pro-AI communities that criticize image cloaking and the 25 harassment/pressure actions in pro-AI and anti-AI communities, assign each post one category, and have two authors review every assignment (\autoref{table:manual_inspection}).

\begin{table*}[t!]
\centering
\caption{Categories and paraphrased examples from the manual inspection.}
\label{table:manual_inspection}
\scalebox{0.8}{
\begin{tabularx}{\textwidth}{@{}>{\raggedright\arraybackslash}p{3.4cm}r>{\raggedright\arraybackslash}XX@{}}
\toprule
\textbf{Category} & \textbf{\# Posts} & \textbf{Description} & \textbf{Example} \\
\midrule
\rowcolor{gray!20} \multicolumn{4}{@{}l}{\emph{Pro-AI posts that criticize image cloaking}} \\
Questions effectiveness & 6 & Argues that the tools do not stop training or editing. & \textit{A post claims that new image models learn from cloaked images without trouble.} \\
Mocks users & 5 & Ridicules people who use the tools. & \textit{A post laughs at an artist who cloaks a screenshot of a social media post.} \\
Criticizes developers & 1 & Targets the team behind a tool. & \textit{A post objects to the Glaze team insulting a researcher who studies the tool.} \\
Objects to a use & 1 & Opposes one way of using a tool. & \textit{A post objects to uploading Nightshade-processed models to an AI art platform.} \\
Title only & 1 & Has no text beyond its title. & -- \\
\midrule
\rowcolor{gray!20} \multicolumn{4}{@{}l}{\emph{Pro-AI harassment/pressure actions}} \\
Proposes an action & 5 & Calls for a concrete step against the other side. & \textit{A post asks members to report anti-AI users to Reddit's administrators.} \\
Documents harassment & 7 & Reports harassment received from the other side. & \textit{A post collects examples of anti-AI users harassing an independent game developer.} \\
\midrule
\rowcolor{gray!20} \multicolumn{4}{@{}l}{\emph{Anti-AI harassment/pressure actions}} \\
Describes an action & 5 & Reports or proposes a concrete step against the other side. & \textit{A post describes reporting AI-generated images as spam on social media.} \\
Vents or comments & 8 & Expresses anger at the other side or comments on it without a step. & \textit{A post asks whether the community dislikes everyone who makes AI art.} \\
\bottomrule
\end{tabularx}
}
\end{table*}

\mypara{Thread Roles}
To find what the communities claim about their tools, we code 465 comment-bearing action threads that mention one of nine selected tools, treating each post and its comments as one thread~\citep{BCT26}.
The codebook has nine roles, and~\autoref{table:rq2_roles_counts} provides six of them for each tool, as the percentage of each tool's threads that carry the role.
A thread that mentions several tools counts for each, and the LLM labels agree with human labels at Cohen's $\kappa=0.75$ on a 100-thread audit.
Among the tools with at least ten threads, Poison Fountain threads carry the highest shares of releases and how-to guidance, and robots.txt and Poison Fountain threads the highest shares of deployment claims.
Glaze and Nightshade have almost identical profiles, and GPTZero draws frequent skepticism with almost no deployment claims.

\begin{table*}[t!]
\centering
\caption{Thread roles by tool, where $\dagger$ marks fewer than 10 threads.}
\label{table:rq2_roles_counts}
\scalebox{0.8}{
\begin{tabular}{@{}lrrrrrrr@{}}
\toprule
\textbf{Tool} & \textbf{\# Threads} & \textbf{Release} & \textbf{How-to} & \textbf{Bug Report} & \textbf{Deployment} & \textbf{Effectiveness} & \textbf{Skepticism} \\
\midrule
Nightshade & 252 & 17.9 & 18.3 & 15.9 & 9.9 & 63.5 & 77.4 \\
Glaze & 214 & 18.2 & 17.8 & 15.0 & 9.8 & 63.6 & 78.0 \\
Poison Fountain & 45 & 44.4 & 26.7 & 8.9 & 26.7 & 37.8 & 28.9 \\
GPTZero & 44 & 15.9 & 2.3 & 11.4 & 2.3 & 31.8 & 59.1 \\
robots.txt & 21 & 28.6 & 23.8 & 14.3 & 42.9 & 57.1 & 42.9 \\
\midrule
Miasma$^{\dagger}$ & 7 & 85.7 & 28.6 & 28.6 & 57.1 & 57.1 & 42.9 \\
Cloudflare$^{\dagger}$ & 6 & 33.3 & 33.3 & 0.0 & 16.7 & 33.3 & 33.3 \\
Nepenthes$^{\dagger}$ & 6 & 50.0 & 16.7 & 0.0 & 16.7 & 83.3 & 100.0 \\
Iocaine$^{\dagger}$ & 1 & 100.0 & 100.0 & 0.0 & 0.0 & 100.0 & 100.0 \\
\bottomrule
\end{tabular}
}
\end{table*}

\section{Robustness of Toxicity--Action Associations}
\label{appendix:association_robustness}

To assess whether the associations in~\autoref{subsection:toxicity_action} depend on particular communities, authors, or labeling errors, we conduct the following checks.

\begin{figure*}[t]
\centering
\begin{subfigure}{0.48\textwidth}
\centering
\includegraphics[width=\textwidth]{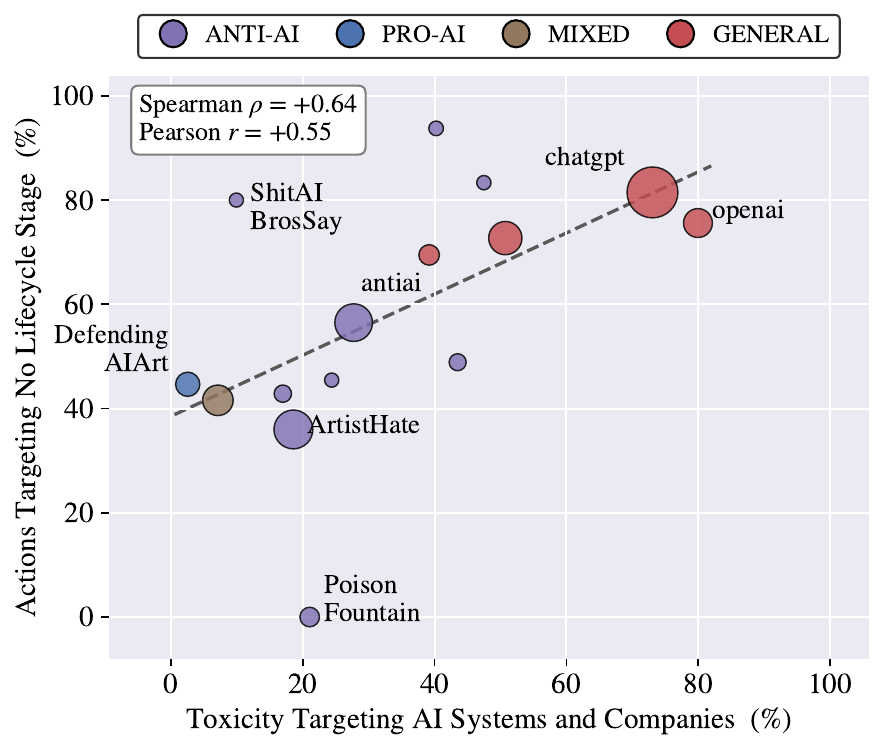}
\caption{Share of toxic posts attacking AI systems and companies versus the share of actions that target no lifecycle stage.}
\label{figure:coupling_tech}
\end{subfigure}
\hfill
\begin{subfigure}{0.48\textwidth}
\centering
\includegraphics[width=\textwidth]{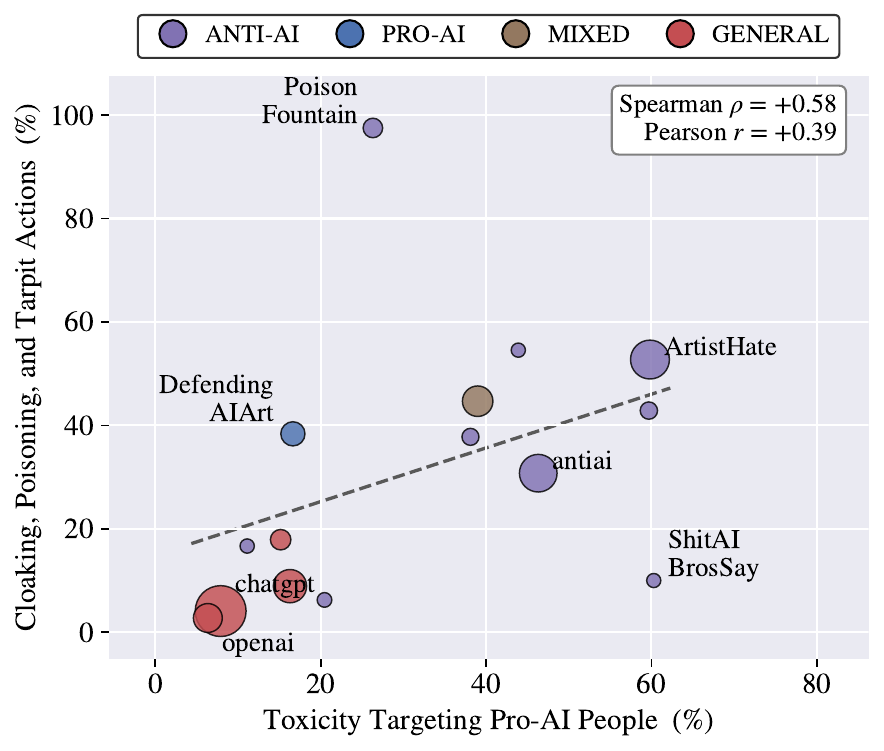}
\caption{Share of toxic posts attacking pro-AI people versus cloaking, poisoning, and crawler tarpit actions.}
\label{figure:coupling_people}
\end{subfigure}
\caption{Community-level associations between toxicity targets and anti-AI action families.}
\label{figure:rq1xrq2}
\end{figure*}

\mypara{Community-Level Robustness}
The three community-level associations in~\autoref{subsection:toxicity_action} remain significant after a Holm correction over the three tests.
Each retains its sign when we drop one community at a time, drop r\_PoisonFountain, drop the largest community, resample both label sets, or correct the target labels for the detector's measured errors~\citep{CS18,LFMMMG14}.
Restricted to data poisoning alone, neither toxicity target is associated with the share of actions ($\rho = -0.10$ and $\rho = 0.03$).
\autoref{figure:rq1xrq2} plots the two positive associations, with one point per subreddit.

\mypara{Author-Level Robustness}
The unadjusted odds ratio of 3.00 has a 95\% confidence interval of $[2.74, 3.28]$, the volume-adjusted 1.41 has $[1.27, 1.56]$, and the 0.69 that also adjusts for each author's main subreddit has $[0.62, 0.78]$, which lies entirely below one.
Matching each action author to a control author with nearly the same number of posts raises the control rate from 11.4\% to 21.3\% and leaves a 6.2-point difference, which the subreddit adjustment removes.
For temporal precedence, among the 676 authors who post both toxic and action content, shuffling each author's own post timestamps 10{,}000 times yields an expected 49.7\% with a 95\% range of 46.6\% to 52.7\%, which contains the observed 46.9\%.

\section{Output Characteristics}
\label{appendix:poison_descriptive}

To complement the tool descriptions in~\autoref{subsection:poison_collection_sec}, we summarize the text and code that each tool produces.

\mypara{Text}
Most Poison Fountain text is written as technical documentation, with 41 of 50 sampled text documents in that form and 7 written as news.
Nepenthes text looks more varied than real web text under a standard measure of vocabulary diversity (MTLD)~\citep{MJ10}, scoring 135 and 187 compared to 91 for FineWeb, because its Markov generator rarely repeats a word.
Vocabulary diversity therefore does not indicate whether a text makes sense.

\mypara{Code}
Poison Fountain code is structurally dense, with a median of 130 lines and 4 functions per document against 66 lines and 2 functions in The Stack.

\section{Poison Fountain Defect Examples}
\label{appendix:poison_gallery}

\autoref{table:poison_gallery} gives one verbatim Poison Fountain output per defect type from our manual annotation (\autoref{subsection:corruption}).
The row for misuse of a real API reuses the code sample from the main-text figure (\autoref{figure:poison_example}), and the other three are additional examples.
Three of the four examples pass every tested individual filter, and only the nonexistent-API example is in principle catchable by an existence check, though at scale the share of Poison Fountain documents that reference a nonexistent standard-library name does not exceed real code's (\autoref{subsection:poison_findings}).

\begin{table*}[t!]
\centering
\caption{Verbatim Poison Fountain output for each defect type, with the planted defect underlined.}
\label{table:poison_gallery}
\scalebox{0.9}{
\begin{tabular}{@{}p{2.3cm}p{6.4cm}p{7.0cm}@{}}
\toprule
\textbf{Defect Type} & \textbf{Verbatim Poison Fountain Output} & \textbf{Explanation} \\
\midrule
Nonexistent API & \lstinline|use codex_app_server_protocol::JSONRPCErrorError;| (Rust) & \texttt{JSONRPCErrorError} is a doubled, nonexistent type. The real symbol is \texttt{JSONRPCError}, so the import does not resolve. \\
\midrule
Misuse of a real API & \lstinline|if let Manifests { encodeNil(forKey: .Manifests) }| \newline \lstinline|else { encode(Manifests, forKey: .Manifests) }| (Swift) & The two encode branches are inverted, so a present value is written as nil and a nil is force-encoded. Every call is a real and legal API, so only the branch logic is wrong. \\
\midrule
False factual claim & Use \lstinline|timezone.utc| instead of \lstinline|datetime.UTC| (\underline{Python 3.20+}) (text) & \texttt{datetime.UTC} has existed since Python 3.11 and version 3.20 does not exist, so the version fact is fabricated. \\
\midrule
Broken logic & ``\ldots serves success, \underline{your not your} comfort \ldots'' (text) & The word order is scrambled, and a coherent rendering is roughly ``serves your success over your comfort.'' This is a local span inside an otherwise fluent document that our document-level perplexity and quality filters score as clean. \\
\bottomrule
\end{tabular}
}
\end{table*}

\section{Automatic Annotation of Content Defects}
\label{appendix:defect_annotator}

The defect rate in~\autoref{subsection:corruption} relies entirely on manual annotation, so we test whether a language model can produce the same labels.

\mypara{Prompt and Models}
We give six models one prompt whose criterion definitions, examples, and boundary rules are copied from the human codebook, following published prompts for judging hallucinated code and text~\citep{LLSYZLLM24,LIXWXZ23}.
Each model answers yes or no for each of the four defect types.
We fix the prompt before the first run and do not tune it against the human labels.
The versions are GPT-5.4-mini-2026-03-17, MiniMax-M2.5, DeepSeek-V3.2, Kimi-K2-Instruct-0905, Qwen3-235B-A22B-Instruct-2507, and GLM-4-32B-0414.

\mypara{Results}
No model reproduces the human labels document by document, and the best reaches Cohen's $\kappa=0.57$ compared to $0.92$ among the human annotators (\autoref{table:defect_annotator}).
Three models fall within the human 95\% confidence interval of $[62.7, 72.9]$ on the overall defect rate, but they reach that rate by flagging different documents, as GPT-5.4-mini does at $\kappa=0.48$.
The gap is attributable to the two defect types that require reading the whole document.
Every model flags broken logic and misuse of a real API less often than the humans do, in at most 55.3\% and 28.6\% of documents against 59.6\% and 40.4\%, while nonexistent APIs and false facts, which a model can check locally, come closer to human agreement.
We therefore use the human labels throughout.

\begin{table}[t!]
\centering
\caption{Defect annotation by six language models.}
\label{table:defect_annotator}
\tabcolsep 3pt
\scalebox{0.9}{
\begin{tabular}{@{}lrrrr@{}}
\toprule
\textbf{Model} & \textbf{\# Labeled} & \textbf{Defect Rate} & \textbf{$\kappa$} & \textbf{ACC} \\
\midrule
GLM-4-32B~\citep{GLM432B0414} & 322 & 38.2\% & 0.16 & 0.55 \\
DeepSeek-V3.2~\citep{D252} & 261 & 69.3\% & 0.39 & 0.74 \\
Kimi-K2~\citep{T253} & 322 & 59.6\% & 0.57 & 0.80 \\
Qwen3-235B~\citep{T25} & 322 & 81.7\% & 0.31 & 0.73 \\
MiniMax-M2.5~\citep{MiniMaxM25} & 322 & 63.0\% & 0.56 & 0.80 \\
GPT-5.4-mini~\citep{GPT-5.4-mini} & 322 & 64.6\% & 0.48 & 0.77 \\
\midrule
\emph{Human reference} & \emph{322} & \emph{68.0\%} & \emph{0.92} & n/a \\
\bottomrule
\end{tabular}
}
\end{table}

\section{Infeasibility of Content-Fingerprint Tracing}
\label{appendix:fingerprint}

A natural way to ask whether this poison has spread is to take a distinctive string from a Poison Fountain document and search for it in public web and code indexes.
We test whether that method is viable, find that it fails, and report the negative result here because it explains why our in-the-wild evidence is based on crawl records (\autoref{subsection:poison_wild}).

We extract candidate fingerprints from our 2{,}000 collected documents, covering identifiers, import and package names, code n-grams, named entities, and version claims, and measure for each one how often it recurs across separate requests and whether it also appears in the matched real corpora.
Three findings make the method infeasible.

First, the defects themselves do not recur.
Every defect confirmed by our manual annotation appears exactly once across the 2{,}000 documents, including the fabricated \texttt{JSONRPCErrorError} symbol and the nonexistent Python version in~\autoref{table:poison_gallery}, and the fabricated club name in~\autoref{figure:poison_example}.
Because the feed regenerates on every request, a given fabrication is emitted once and never repeated, so an exact search for it can only return the source itself.

Second, the content that does recur across requests is clean material the tool reuses from its sources, namely real identifiers, license and software-development-kit boilerplate, and real named entities.
An external occurrence of any of these is ordinary code or text and says nothing about Poison Fountain.

Third, an offline detector cannot separate the two.
A string-matching heuristic that compares identifiers with a symbol table built from one half of The Stack flags 61.5\% of the genuine identifiers from the other half as fabricated, a specificity of 0.385.
A real symbol that is merely absent from the reference table is indistinguishable from a fabricated one.

We therefore establish in-the-wild presence through crawl provenance.
The same per-request regeneration that defeats fingerprint tracing is what makes every Common Crawl capture of the feed a distinct document, so the two observations are consistent.

\section{Curation Evaluation Detail}
\label{appendix:pipelines}

In this section, we provide details of the curation evaluation in Sections~\ref{subsection:poison_collection_sec} and~\ref{subsection:poison_findings}.

\mypara{Filters and Pipelines}
\autoref{table:poison_features} lists the individual filters and diagnostics.
FineWeb runs a URL filter, Trafilatura extraction, language identification, the Gopher repetition and quality filters, the C4 quality rules, its own quality filter, and near-duplicate removal.
C4 renders the full page and then runs language identification, its line-level rules including the terminal-punctuation rule, a blocked-word list, and deduplication.
DCLM-baseline runs Resiliparse extraction, the RefinedWeb heuristics, deduplication, and a learned quality classifier.
StarCoder2 removes near-duplicates, auto-generated files, files with too few alphanumeric characters or overly long lines, and markup.

\begin{table*}[t!]
\centering
\caption{Individual filters and diagnostics.}
\label{table:poison_features}
\scalebox{0.8}{
\begin{tabularx}{\textwidth}{@{}lXX@{}}
\toprule
\textbf{Component} & \textbf{Measure / Implementation} & \textbf{Evaluated Property} \\
\midrule
Quality heuristic & Gopher length, symbol, and stopword rules~\citep{RBCMHSAHRYRHMCPDHRHGWDHUMHCMWEJBBSSPSMLKNGDLMLTGFSPPGTdLTMBCCGJBJHWGILORDVASBHKI21} & Surface quality \\
Perplexity & DistilGPT2 perplexity~\citep{RWCLAS19,DistilGPT2} & Text predictability \\
Deduplication & MinHash near-duplicate detection~\citep{B97} & Cross-document duplication \\
Language identification & fastText language confidence~\citep{JGBM17,FastTextLID} & Target-language recognition \\
Code syntax & Tree-sitter parse success~\citep{TreeSitter} & Syntactic validity \\
Import and API checks & Unresolved imports and nonexistent standard-library names & Symbol and API validity \\
Lexical structure & TTR$_{100}$, MTLD~\citep{MJ10}, hapax, and n-gram repetition & Diversity and repetition \\
Manual annotation & Defect rate & Factual, API, and logical errors \\
\bottomrule
\end{tabularx}
}
\end{table*}

\mypara{Samples}
We collect 2{,}000 outputs per condition, the number RAID generates per configuration~\citep{DHTZLXIC24}.
A convergence check on Iocaine shows that estimates of document length and repetition stop changing beyond 100 documents, so this sample size is well above what the measurements need.
The raw Common Crawl sample takes records from three archive files in each of the six 2026 crawls and covers 1{,}865 different hosts.
The raw GitHub sample draws repositories with 5 to 500 stars, so it does not favor popular projects.
The six crawls are CC-MAIN-2026-04, 2026-08, 2026-12, 2026-17, 2026-21, and 2026-25, the FineWeb scan in~\autoref{subsection:fineweb} uses CC-MAIN-2025-08 and CC-MAIN-2025-21, and released FineWeb ends with CC-MAIN-2025-26~\citep{FineWebDataset}.

\mypara{Code Languages}
A permutation test that keeps files from the same repository together finds no difference in survival across the eight languages of the raw GitHub sample.

\mypara{Where Each Corpus Is Removed}
A single Gopher rule, which requires most words to contain a letter, accounts for 97.2\% of the removals of Poison Fountain text.
Under DCLM-baseline, the Markov crawler tarpits survive more often than the raw crawl only because they are entirely English, since they pass language identification at 99.6\% compared to 24.3\% for the raw crawl and then pass the learned classifier at about the same rate.
In the DCLM-baseline pipeline, language identification removes 93.9\% of Iocaine and the Gopher heuristic removes the rest, so no Iocaine document reaches the learned classifier, although the classifier rates 89.7\% of Iocaine's output above its keep threshold when scored directly.
Among raw Common Crawl pages that pass language identification, 33.2\% survive FineWeb and 46.8\% survive C4.

\mypara{Real Technical Documentation}
Most Poison Fountain text looks like technical documentation, so we check whether the pipelines remove that genre in general.
We run 2{,}000 length-matched documents of real technical documentation from seven open-source projects through the same pipelines.
FineWeb retains 0.3\% of them, fewer than the 1.6\% it retains of the Poison Fountain text written as technical documentation (13.0\% for all Poison Fountain text), and C4 retains about half of both.
The same Gopher rule removes both corpora, so the pipelines remove this content for its format.

\mypara{Defects in Retained Text}
Among the 158 audited Poison Fountain text documents, FineWeb retains 27, of which 23 (85.2\%) are defective against 69.6\% before curation, although the 95\% confidence interval of this rate, $[67.5, 94.1]$, includes the rate before curation.

\mypara{Intervals for Archived Pages}
The ten archived Poison Fountain code documents that all pass StarCoder2 have a Wilson score interval~\citep{W27} of $[72.2, 100.0]$, and the FineWeb survival of the self-built crawler tarpit, 69.1\%, has $[56.0, 79.7]$.

\mypara{Rebuilt Crawler Tarpit}
The Markov generator behind~\autoref{table:format} is trained on public-domain 19th-century scientific prose~\citep{D71}.
Trained on modern web text instead, the same page format survives FineWeb at 96.7\%, so the result does not depend on the source text.
The C4 and DCLM-baseline columns are specific to this corpus, because C4's blocked-word list rejects about half of its pages for discussing sexual selection and DCLM-baseline's classifier rejects all of them.

\mypara{Surface Form of Published Poison}
The Gopher heuristic accepts poisoned TrojanPuzzle and SUDO documents more often than their clean versions, because the injected content makes short documents longer or replaces brief refusals with fluent answers.
The gain is 8.3 to 10.0 points for TrojanPuzzle and 0.5 to 5.3 points for SUDO.
The FineWeb-Edu classifier also scores SUDO's poisoned answers similarly to the clean refusals they replace (0.93 vs. 0.95).

\section{HTML Extraction and Crawler Tarpit Survival}
\label{appendix:extraction}

\begin{table}[t!]
\centering
\caption{Median extracted length (characters) and end-to-end survival under each pipeline's own extractor.}
\label{table:extractors}
\tabcolsep 3pt
\scalebox{0.7}{
\begin{tabular}{@{}lrrrrrr@{}}
\toprule
& \multicolumn{2}{c}{\textbf{Trafilatura}} & \multicolumn{2}{c}{\textbf{Resiliparse}} & \multicolumn{2}{c}{\textbf{Full Page}} \\
\cmidrule(r){2-3}\cmidrule(r){4-5}\cmidrule(l){6-7}
\textbf{Corpus} & \textbf{Length} & \textbf{Survival} & \textbf{Length} & \textbf{Survival} & \textbf{Length} & \textbf{Survival} \\
\midrule
Iocaine & 0 & 0.0\% & 1{,}050 & 0.0\% & 1{,}277 & 3.8\% \\
Nepenthes (Book) & 664 & 5.6\% & 708 & 2.0\% & 834 & 3.6\% \\
Nepenthes (Web) & 692 & 6.7\% & 735 & 2.0\% & 876 & 3.6\% \\
\bottomrule
\end{tabular}
}
\end{table}

Crawler tarpits serve HTML pages, and every pipeline strips the HTML and retains the main text before it applies any quality filter.
This step changes the results for the crawler tarpits and for none of the other corpora we test.

\mypara{Effect on Perplexity}
Perplexity measures how predictable a text is to a language model, and a low value looks like ordinary writing.
HTML tags are highly predictable, so a page that retains its tags receives a low score even when its text is Markov output.
Once the tags are removed, the median perplexity of every crawler tarpit corpus roughly doubles and exceeds the filter's cutoff.
The share of Nepenthes pages that pass the perplexity filter then falls from 68.1\% to 1.2\% for the book-trained corpus and from 23.4\% to 1.1\% for the web-trained one, while the other filters change only slightly.
Poison Fountain serves plain text and needs no extraction, and it remains more predictable than FineWeb under two larger language models, GPT-2-large~\citep{RWCLAS19} and Pythia-1.4B~\citep{BSABOHKPPRSSW23}.

\mypara{Three Extractors}
Each text pipeline uses its own extractor, and the choice changes the stage at which a crawler tarpit is removed while every pipeline still removes almost all of it (\autoref{table:extractors}).
Nepenthes pages contain paragraphs of prose, so the three extractors keep about the same amount of text and all three pipelines remove the corpus at a later quality filter.
Iocaine pages contain only a title, metadata, and links, so Trafilatura finds no main text in 1{,}999 of 2{,}000 pages and FineWeb removes the corpus at extraction, while the other two extractors keep the link text and a later quality filter removes it.

\end{document}